\pdfoutput=1
\documentclass[12pt]{emulateapj}
\usepackage{lscape}
\usepackage{subfigure}
\usepackage{graphicx}
\usepackage{multirow}
\usepackage{latexsym}
\usepackage{amssymb}
\usepackage{mathtools}
\usepackage{epsf}
\usepackage{threeparttable}
\usepackage{amsmath}

\newcommand{\kms}{km s$^{-1}$ }
\newcommand{\kmsn}{km s$^{-1}$}
\newcommand{\Msu}{$M_{\odot}$ }
\newcommand{\Msun}{$M_{\odot}$}

\newcommand{\hi}{{\rm H\,}{{\sc i}}}
 
\newcommand{\his}{{\rm H\,}{{\sc i }}}

\newcommand{\COT}{$^{12}$CO}
\newcommand{\COTs}{$^{12}$CO }

\newcommand{\logten}{\ensuremath{\log_{10}}}

\begin{document}	\title{Dust Abundance Variations in the Magellanic Clouds: Probing the Lifecycle of Metals with All-Sky Surveys}

\author{Julia Roman-Duval\altaffilmark{1}, Caroline Bot\altaffilmark{2}, Jeremy Chastenet\altaffilmark{1,2}, Karl Gordon\altaffilmark{1}}
\altaffiltext{1}{Space Telescope Science Institute, 3700 San Martin Drive, Baltimore, MD 21218; duval@stsci.edu}
\altaffiltext{2}{Universit\'{e} de Strasbourg, CNRS, Observatoire astronomique de Strasbourg, UMR 7550, F-67000 Strasbourg, France}

\begin{abstract}
Observations and modeling suggest that the dust abundance (gas-to-dust ratio, G/D) depends on (surface) density. The variations of the G/D provide constraints on the timescales for the different processes involved in the lifecycle of metals in galaxies. Recent G/D measurements based on Herschel data suggest a factor 5---10 decrease in the dust abundance between the dense and diffuse interstellar medium (ISM) in the Magellanic Clouds. However, the relative nature of the Herschel measurements precludes definitive conclusions on the magnitude of those variations. We investigate the variations of the dust abundance in the LMC and SMC using all-sky far-infrared surveys, which do not suffer from the limitations of Herschel on their zero-point calibration. We stack the dust spectral energy distribution (SED) at 100, 350, 550, and 850 microns from IRAS and Planck in intervals of gas surface density, model the stacked SEDs to derive the dust surface density, and constrain the relation between G/D and gas surface density in the range 10---100 \Msu pc$^{-2}$ on $\sim$ 80 pc scales. We find that G/D decreases by factors of 3 (from 1500 to 500) in the LMC and 7 (from 1.5$\times 10^4$ to 2000) in the SMC between the diffuse and dense ISM. The surface density dependence of G/D is consistent with elemental depletions and with simple modeling of the accretion of gas-phase metals onto dust grains. This result has important implications for the sub-grid modeling of galaxy evolution, and for the calibration of dust-based gas mass estimates, both locally and at high-redshift. 

\end{abstract}

\maketitle

\section{Introduction} \label{introduction}

\begin{figure*}
\centering 
\includegraphics[width=8cm]{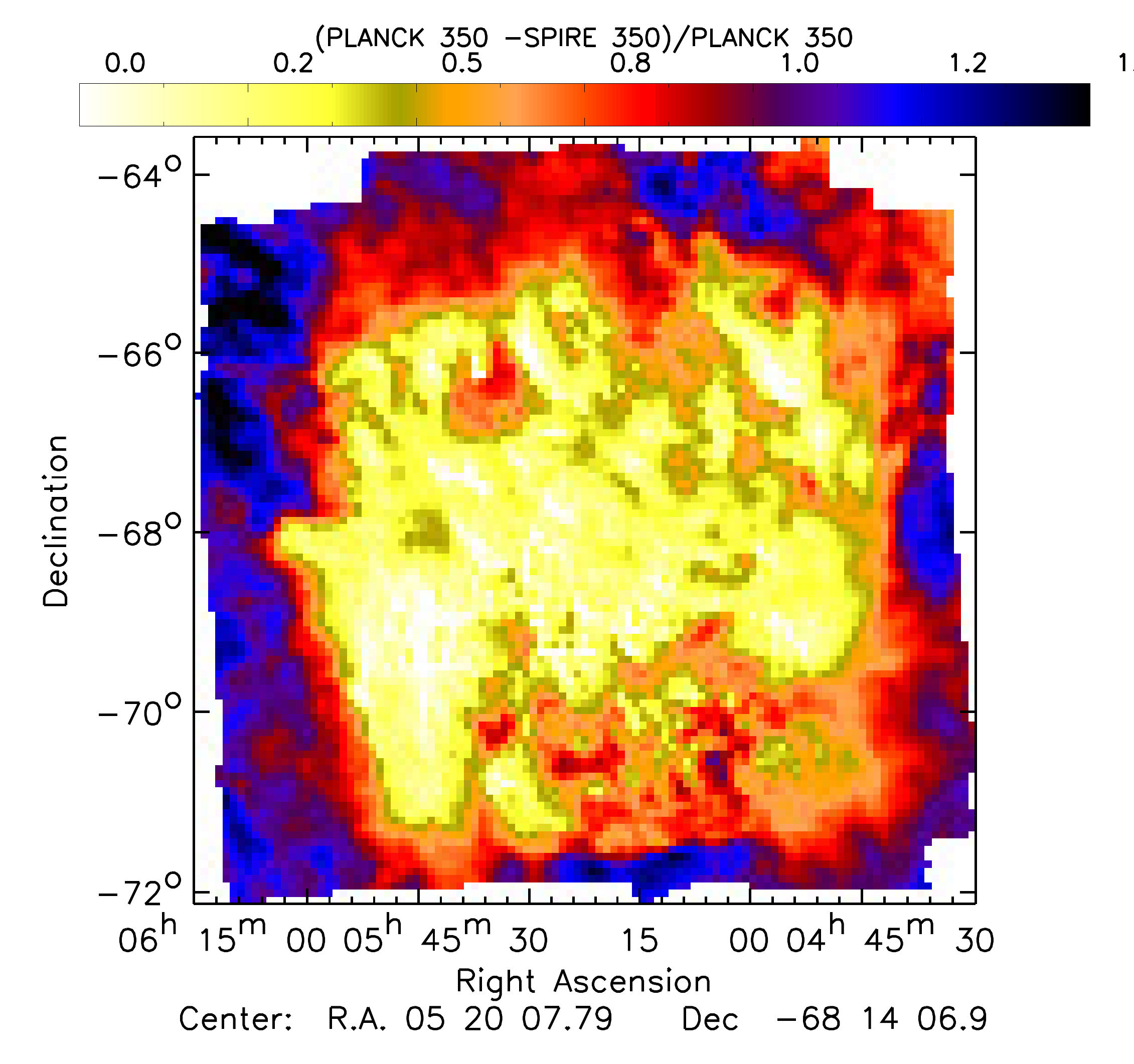}
\includegraphics[width=8cm]{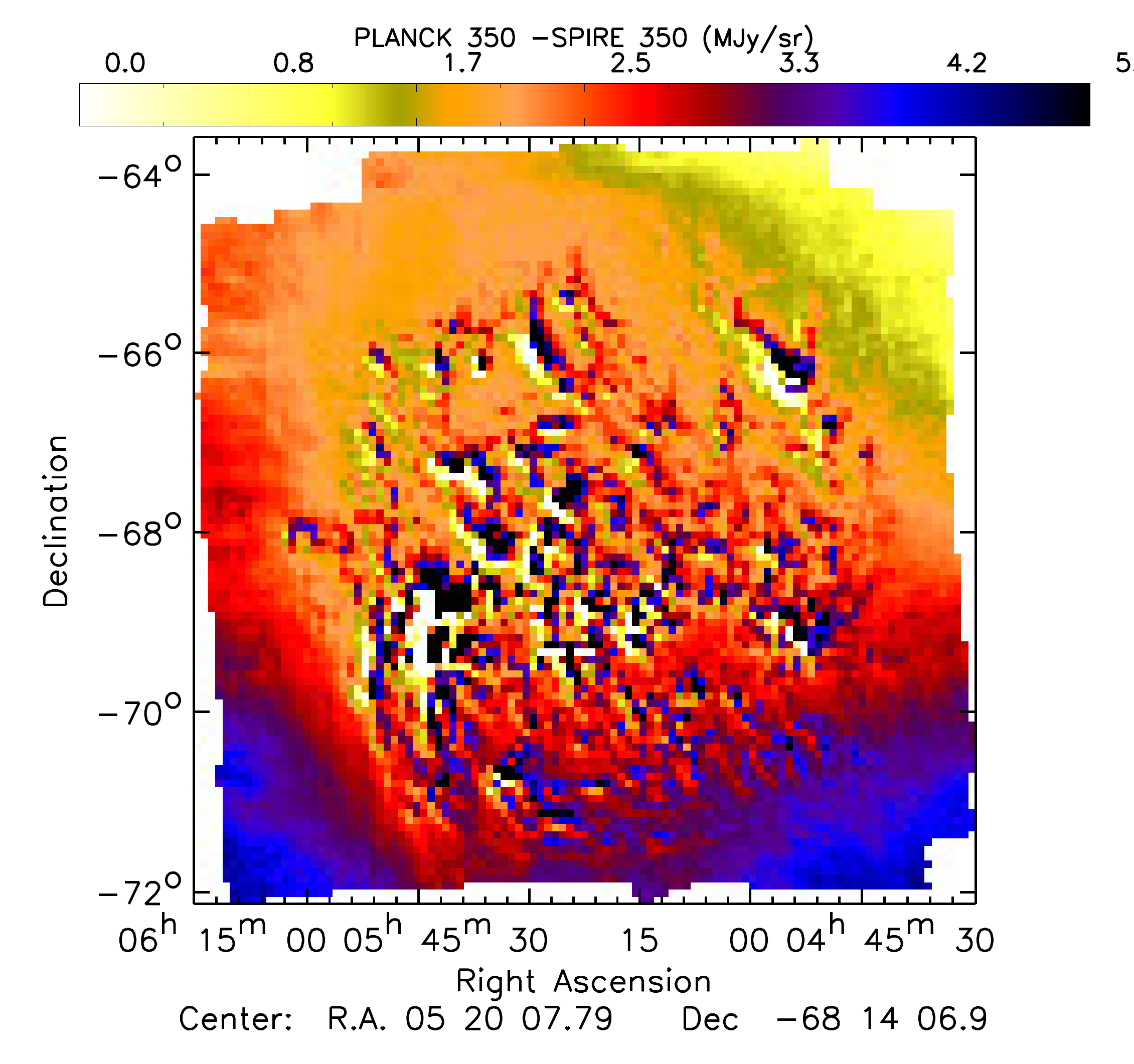}
\caption{Image of the fractional (left) and absolute (right) difference between the 350 $\mu$m band of PLANCK and {\it Herschel}/SPIRE in the LMC, where little background is available to evaluate instrumental drifts and 1/f noise. The fractional and absolute differences are computed as (PLANCK - SPIRE)/PLANCK and (PLANCK-SPIRE), respectively.}
\label{planck_spire_dif_image}
\end{figure*}

\indent Dust grains absorb stellar light in the UV-optical and re-emit it in the FIR, which represents 30---50\% of the output of a galaxy. Therefore, our ability to interpret observations of galaxies and
trace their stellar, dust, and gas content over cosmic time over the entire spectral range critically relies on our understanding of how the dust abundance and properties vary with environment; this
in turn requires us to understand the processes responsible for dust formation, destruction, and transport and their associated timescales. Addressing the question of dust growth and destruction
requires large-scale maps of the dust (in the far-infrared, FIR) and gas (HI and CO) content of galaxies resolved on kpc-scales and over a range of metallicities. Subsequent gas-to-dust ratio and dust property measurements can then be compared to phenomenological, ÒresolvedÓ models of the chemical and dust evolution of entire galaxies. These models have advanced to the point of needing more
detailed, environment-dependent observations of dust abundances and properties. Since variations in the dust abundance and properties cannot be observationally characterized in unresolved
galaxies, we must therefore characterize and understand the micro-astrophysical processes responsible for dust evolution in the ISM in nearby galaxies, and apply our knowledge to the
more distant universe.\\
\indent We have been investigating dust evolution in the Magellanic Clouds (LMC and SMC) by measuring their dust and gas contents using over 1400 hours of space observatory time with Spitzer \citep[Surveying the Agents of Galaxy Evolution, SAGE survey][]{meixner2006, gordon2011} and Herschel \citep[HERschel Inventory of The Agents of Galaxy Evolution, HERITAGE survey][]{meixner2013}, in addition to ground-based measurements of CO \citep{mizuno2001, mizuno2001_smc, wong2011}, HI \citep{stanimirovic1999, kim2003}, and H$\alpha$ \citep{gaustad2001}. Thanks to these observations, we now have dust surface density and temperature maps in the LMC and SMC \citep{gordon2014}. We have developed sophisticated tools to extract accurate dust parameters (surface density, temperature etc.) from the images using probabilistic techniques \citep[see ][for details]{gordon2014}. In particular, systematics that can potentially affect dust and gas measurements in the low signal-to-noise, low surface density regime were carefully addressed and removed. For instance, covariance between different bands was accounted for; the dust emissivity was self-calibrated on Milky Way observations; and a foreground subtraction was done to remove the structured emission due to Milky Way (MW) dust (cirrus) emission. This step was particularly important for the SMC, where foreground Milky Way structures with similar surface brightness to those in the SMC were removed.\\
\indent In \citet{RD2014}, we examined the resolved relation between dust and gas surface densities in the LMC and SMC. We discovered that 1) the hydrogen-to-dust ratio (H/D) in the diffuse atomic ISM is 380$^{+250}_{-130}$ in the LMC and 1200$^{+1600}_{-420}$ in the SMC; 2) the dust abundance (dust-to-hydrogen ratio, D/H) and FIR emissivity increase by a factor 2---3 between the diffuse atomic and the dense molecular ISM; 3) there is a gas pedestal of 10 \Msu pc$^{-2}$ (LMC) and 40 \Msu pc$^{-2}$ (SMC) in the relation between dust and hydrogen gas, implying the presence of a dust-poor ISM component at surface densities barely or not detected by Herschel (below the gas pedestal) located in the outskirts of the galaxies and holes in the ISM.  This dust-poor ISM component would have a D/H 5---10 times lower than in the FIR detected regions and correspond to the junction between the ISM and the circum-galactic medium (CGM). This hypothesis explaining the non-zero intercept of the dust-gas relation is supported by a stacking analysis of the pixels of the FIR maps where the brightness is below the sensitivity cut and a dust surface density cannot be derived in \citet{gordon2014}. The dust mass corresponding to this FIR-faint component is (5.9$\pm$3.6) $\times$ 10$^4$ \Msu in the LMC and (1.6$\pm$1.3) $\times$ 10$^4$ \Msu in the SMC.  The corresponding \his masses are 1.6 $\times$ 10$^8$  \Msu and 2.0 $\times$ 10$^8$ \Msu in the LMC and SMC, leading to global H/D of $\sim$2800 in the LMC and $\sim$12,000 in the SMC, or 5---10 times higher than in the diffuse atomic ISM detected in the FIR. \\
\indent While this tantalizing result could have important implications for the chemical evolution of the LMC and SMC, and galaxies in general, its robustness is limited by the inherently relative nature of the Herschel (and Spitzer) observations. Indeed, in order to remove 1/f noise (striping) and instrumental drifts from individual scans, one has to make assumptions about the level and structure of the emission in the periphery of the observed galaxies. For observations of galaxies relatively small on the sky, observations usually include a suitable ÒbackgroundÓ area far from the center of the galaxy where zero-emission can reasonably be assumed. For close-by galaxies however, this turns out to be problematic since the observations usually do not extend far enough. As a result, although there is still a significant level of astrophysical signal at the periphery of the images, this low-level emission is zeroed out in the maps. This is illustrated in Figure \ref{planck_spire_dif_image}, which shows a difference image between the 350 $\mu$m bands of Planck and Herschel/SPIRE at 5' resolution. To circumvent this issue in \citet{RD2014}, we performed a background subtraction on the \his 21 cm images to put them on the same pedestal as the dust maps. While this provides a reasonable data set to evaluate variations in the dust abundance, an absolute measurement of the gas-to-dust ratio is not possible. Additionally, issues remain due to the different background assumptions between the scans and cross-scans in the Herschel images. \\
\indent In this paper, we use Planck and IRAS all-sky observations, which are not affected by this limitation on the zero-point calibration and cover the peak of the dust SED at 100, 350, 550, and 850 $\mu$m, to quantify the spatial variations of the dust abundance, from the very diffuse ISM in the outskirts of the LMC and SMC, to the dense ISM where CO emission is detected.  We separate the ISM in intervals of HI column density, and stack the dust emission in each gas surface density bin. We then fit a dust model to the resulting stacked SED to determine the mean dust surface density, temperature, and spectral emissivity index in each bin. We then compare the resulting trend in D/G vs $\Sigma_g$ to resolved chemical evolution models.  In addition to improvements on the zero-points, this stacking technique allows us to probe the dust content of lower surface density regions compared to \citet{RD2014}, and to measure the variations of G/D as a function of surface density.\\
\indent  The paper is organized as follows. In Section \ref{observations_section}, we describe the data sets used in this analysis. In Section \ref{method_section}, we describe the methodology used to subtract the foreground Milky Way cirrus emission, the stacking of the dust emission in intervals of gas surface density, and the dust modeling. In Section \ref{results_section}, we present the main results of this paper, specifically the variations of the dust surface density, gas-to-dust ratio, and other dust parameters with gas surface density. Finally, we discuss the possible origins of the observed trends and their implications in Section \ref{discussion_section}.

\section{Observations} \label{observations_section}

\begin{figure*}
\centering
\includegraphics[height=8cm]{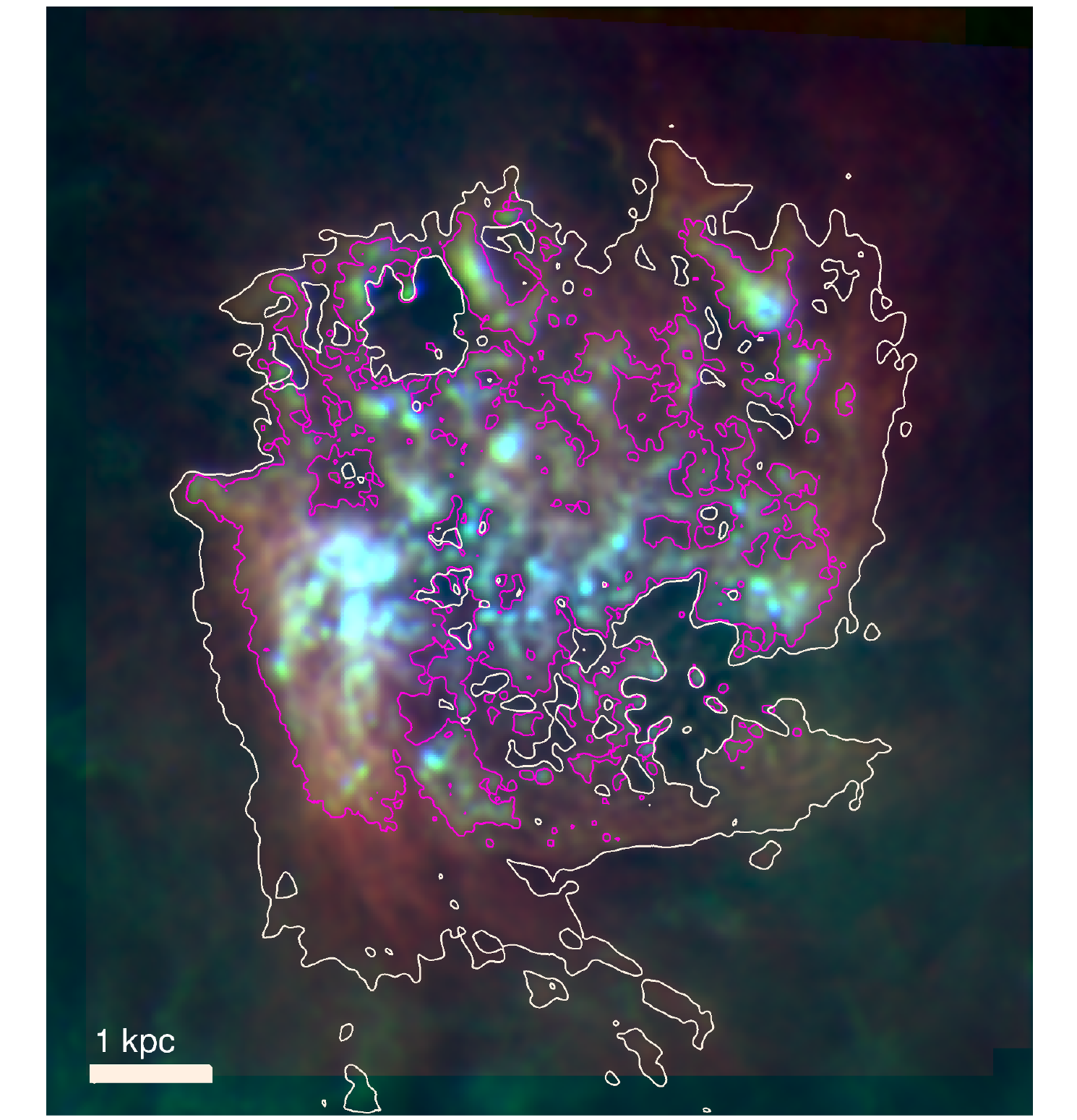}
\includegraphics[height=8cm]{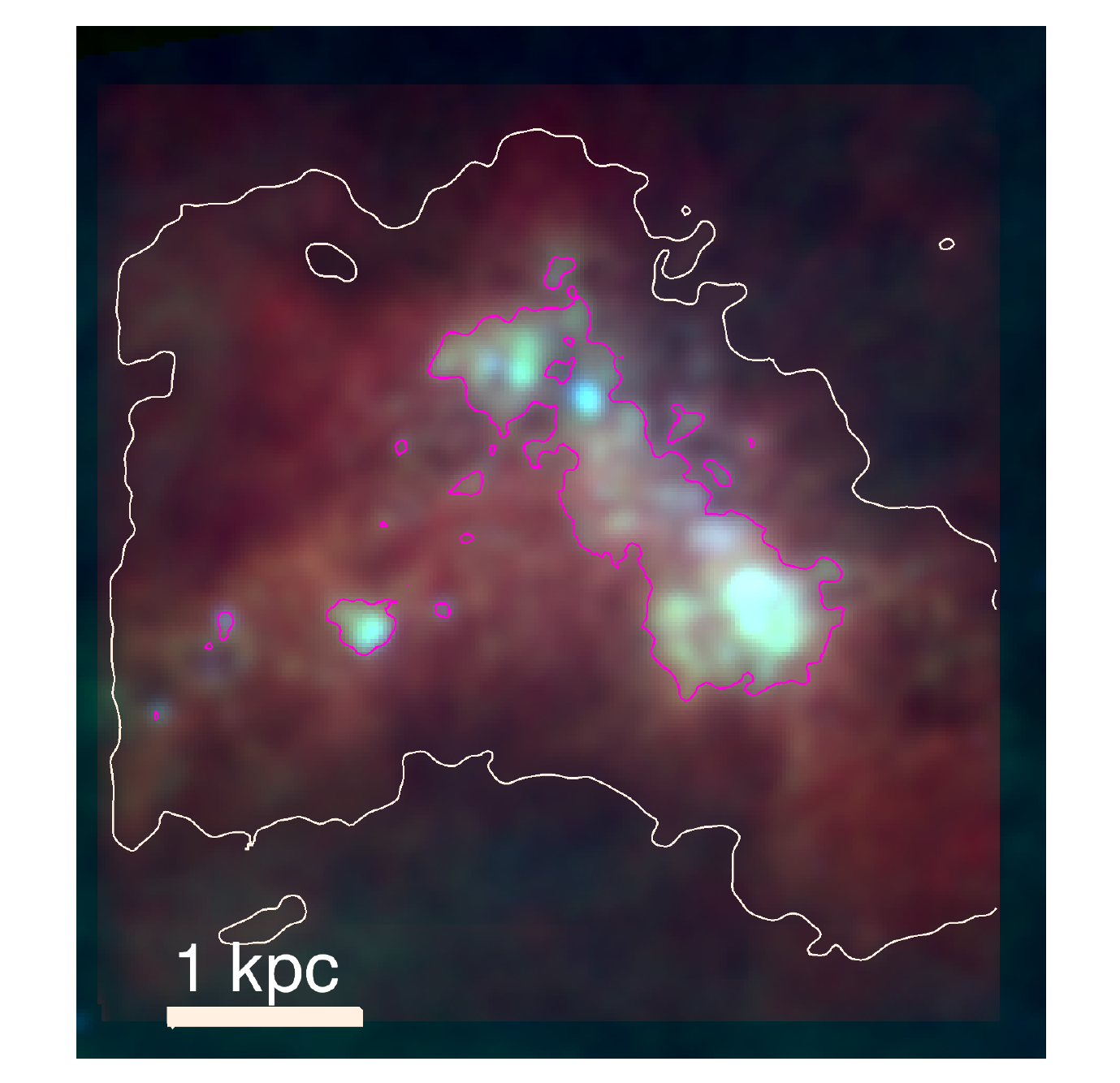}
\caption{Three-color image of the LMC (left) and SMC (right) in dust and gas. The red, green, and blue colors correspond to the HI surface density, the dust emission in the Planck 350 $\mu$m and in the IRAS 100 $\mu$m bands. The images are shown in square-root scale. For the LMC, the ranges shown are 0---130 MJy/sr for IRAS 100 $\mu$m, 0---40 MJy/sr for PLANCK 350 $\mu$m, and 0---200 \Msu pc$^{-2}$ for the \his surface density. For the SMC, the ranges shown are 0---60 MJy/sr for IRAS 100 $\mu$m, 0---20 MJy/sr for PLANCK 350 $\mu$m, and 0---300 \Msu pc$^{-2}$ for the \his surface density. This Figure, which shows variations in the colors, provides a striking visualization of the varying dust abundance and properties from the bright star-forming regions to the outskirts of the LMC and SMC. The foreground cirrus emission from the Milky Way is clearly visible in the PLANCK 350 $\mu$m band south of the LMC and in the lower left corner of the SMC. The magenta contours indicate the area where a dust surface density was derived in \citet{gordon2014}, while the white contours indicate the area included in this analysis. The stacking analysis allows us to probe the dust content of much lower surface densities.}
\label{figs_3col}
\end{figure*}

\indent The observations used in this analysis are described in this section, and include FIR dust emission observed in IRAS and PLANCK, 21 cm emission from \his gas observed with both single dish and interferometry, and CO emission from molecular gas observed with various ground-based millimeter facilities.  Figure \ref{figs_3col} shows the IRAS 100 $\mu$m (blue), Planck 350 $\mu$m (green), and \his surface density (red) maps as three-color images.

\subsection{Far-infrared dust observations}
\indent This work utilizes all-sky observations with IRAS 100 $\mu$m and PLANCK 350, 550, and 850 $\mu$m. We used the ISSA (IRAS Sky Survey Atlas) version of the IRAS observations, which were downloaded from the Infra-Red Science Archive (IRSA).  The Improved reprocessing of the IRAS Survey (IRIS) is affected by a $\sim$ 30\% gain offset when compared to COBE/DIRBE and Herschel  in the area of the Magellanic Clouds \citep{meixner13}, and we hence chose to use the older ISSA version. 

\indent The Planck maps at 350, 550, and 850 $\mu$m (857 GHz, 545 GHz, 353 Ghz) from the 2nd release \citep{planck2015_viii_cal} were downloaded in HEALPIX format from the Planck archive. They were convolved to 5' resolution,  re-projected in WCS format, and converted to MJy/sr. Cosmic microwave background (CMB) emission and fluctuations were subtracted using the map from \citet{bobin2016}, which is the only CMB map not impainted in regions like the Magellanic Clouds and does not show significant residuals of foreground emission in bright regions. \\

\indent Cosmic infrared background (CIB) emission was subtracted using mean intensity levels of 0.11 MJy/sr, 0.64 MJy/sr, 0.35 MJy/sr, and 0.13MJy/sr at 100 $\mu$m, 350 $\mu$m, 550 $\mu$m, and 850 $\mu$m, respectively \citep{planck2016_intermediate_xlviii, planck2015_viii_cal}.

\subsection{Atomic gas}

\subsubsection{Observations}

\indent The atomic gas surface density is derived from 21 cm line emission maps with 60'' and 98'' resolution in the LMC and SMC. The HI maps combine observations carried out at the Australian Telescope Compact Array (ATCA) and at the Parkes 64 m radio Telescope. The combination of the single dish and interferometric observations are described in \citet[][LMC]{kim03} and \citet[][SMC]{stanimirovic1999}. The \his column density maps are converted from the 21 cm emission under the assumption of an optically thin line \citep[e.g.][]{bernard08}. This assumption may not be correct in regions of high \his column density, where we may systematically underestimate the \his content \citep{dickey00, fukui14}.\\

\subsubsection{Conversion between 21 cm emission and \his surface density}

\indent The \his column density in the ATCA+Parkes survey, $N($\hi$)$ in units of cm$^{-2}$, is converted to a surface density, $\Sigma$(\hi), via $\Sigma$(\hi) $=$ $0.8 \times 10^{-20}$ $N($\hi$)$, where $\Sigma$(\hi) is in \Msu pc$^{-2}$. This conversion does not include the mean molecular weight of Helium. The atomic gas surface density, including Helium, is given by $\Sigma_{g,\mathrm{atomic}}$ $=$ 1.36 $\Sigma$(\hi). \\

\subsubsection{\his maps of Milky Way Foreground}

\indent To estimate MW cirrus FIR emission, we use the atomic gas surface density images in the MW velocity range. For the SMC, these are obtained from the ATCA+Parkes cubes \citep{stanimirovic1999}, integrated over the MW velocity range. For the LMC, the MW velocities in the ATCA+Parkes data \citep{kim03} were never reduced, and we make use of the Parkes only cube instead \citep{staveleysmith03}, integrated over MW velocities.

\subsubsection{Stray-light correction}

\begin{figure*}
\includegraphics[width=8cm]{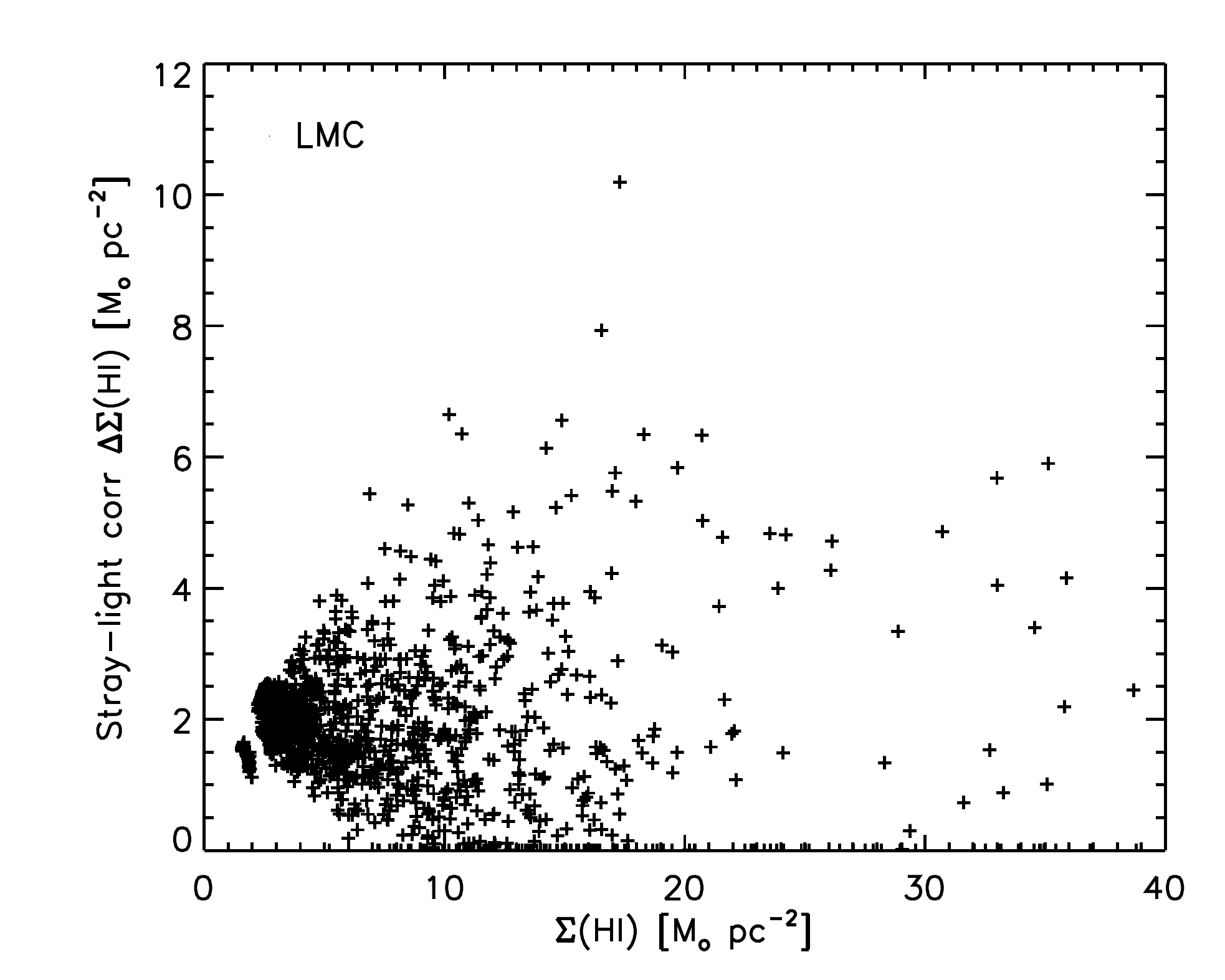}
\includegraphics[width=8cm]{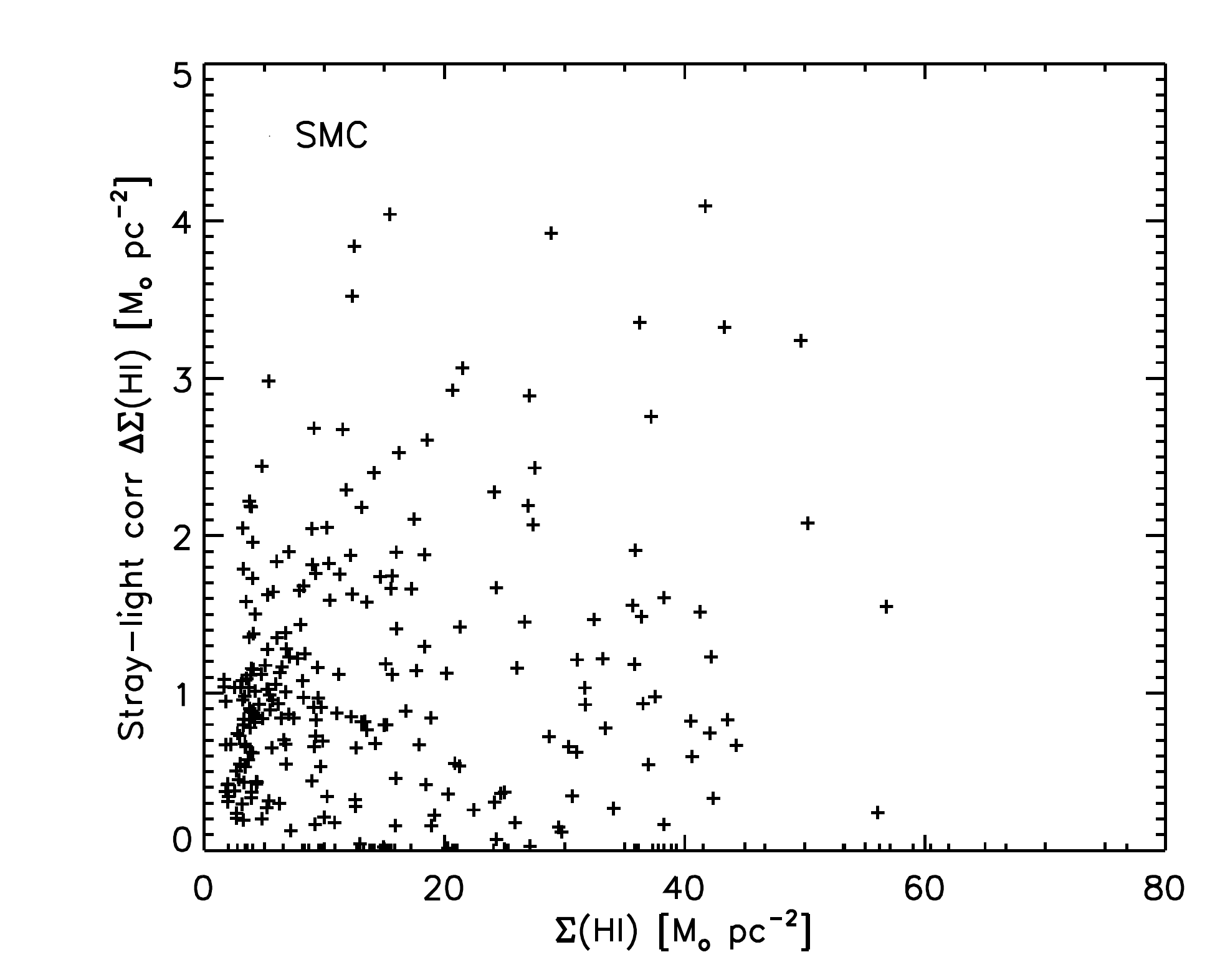}
\caption{Stray-light correction for the ATCA+Parkes  expressed in equivalent \his surface density as a function of \his surface density at the native 16' resolution of the GASS survey.}
\label{plot_stray}
\end{figure*}

\indent The ATCA+Parkes 21 cm data are not corrected for stray-light, which may contribute a low-level of extended emission. To estimate  the contribution of stray-light to the \his surface density measured in the ATCA+Parkes data, we use the corrected GASS survey \citep[16' resolution][]{mclure-griffiths2009, kalberla2015}, which was downloaded from \url{https://www.astro.uni-bonn.de/hisurvey/gass/}. While the GASS survey has too coarse a resolution (16' corresponding to 230 pc in the LMC and 290 pc in the SMC) to be used to derive the G/D trends, it can be used to estimate the contribution of stray-light in the ATCA+Parkes maps. We convolve the ATCA+Parkes maps to GASS resolution using Gaussian kernels and resample the convolved ATCA+Parkes and GASS maps on the same astrometric grid, as described in Section \ref{convolution_section}. We examine the collapsed HI 21 cm spectra in Y-log scale and determined the atomic gas velocity range of the LMC and SMC to be 150---600 \kms and 55-270 \kmsn, respectively. We subsequently extract Milky Way (MW), LMC, and SMC HI  21 cm spectral cubes and integrated intensity images for each galaxy. The cubes were generated in units of K, and so the integrated intensity images are in K km s$^{-1}$. We convert the intensity to a column density as in Equation 2 of \citet{bernard08} via $N($\hi$)$ $=$ 1.82$\times 10^{18}$ $W($\hi$)$, where $N($\hi$)$ is the \his column density in cm$^{-2}$ and $W($\hi$)$ is the 21 cm integrated intensity in K km s$^{-1}$. We compute the difference between the \his column densities derived from the 16' ATCA+Parkes and GASS maps, which corresponds to the contribution of stray-light on scales larger than 16' in the ATCA+Parkes data. The correction, shown in Figure \ref{plot_stray}, amounts to an equivalent \his surface density $<$ 10 \Msu pc$^{-2}$ in the LMC and $<$ 4 \Msu pc$^{-2}$ in the SMC. Finally, we resample the stray-light correction to the astrometric grid of the ATCA+Parkes, which matches that of the other maps (Planck, IRAS, CO), and subtract it from the ATCA+Parkes observations.

\subsection{Molecular gas traced by CO}\label{CO_observations}

\indent The molecular gas component is partially traced by its \COT J = 1-0 emission in the NANTEN survey of the LMC and SMC \citep{mizuno2001_smc} at 2.6$'$ resolution. The NANTEN surveys of the Magellanic Clouds have complete coverage and should capture most of the molecular gas content traceable via CO emission. At this resolution, we assume a CO-to-molecular gas conversion factor $\alpha_{\mathrm{CO}}$ $=$ 6.4 \Msu pc$^{-2}$ (K km s$^{-1})^{-1}$ ($\sim$1.5$\times$ Galactic) in the LMC, and  $\alpha_{\mathrm{CO}}$ $=$ 21 \Msu pc$^{-2}$ (K km s$^{-1})^{-1}$ ($\sim$5$\times$ Galactic) in the SMC \citep{bolatto13}. This conversion factor includes the mean molecular weight of Helium. The molecular gas surface density is then $\Sigma_{g, \mathrm{mol}}$ $=$ $\alpha_{\mathrm{CO}}$ $I_{\mathrm{CO}}$, where $I_{\mathrm{CO}}$ is the \COTs integrated intensity (in K km s$^{-1}$). \\

\subsection{Convolution to common resolution and resampling}\label{convolution_section}

\indent \indent All maps were convolved to the 5' limiting resolution of the Planck observations, corresponding to 70 pc in the LMC \citep[at a distance of 50 kpc][]{} and 90 pc in the SMC \citep[at a distance of 62 kpc][]{}. We used a gaussian kernel, of FWHM equal to the quadratic difference between the final and original resolutions, to perform the convolution. The maps were then resampled onto a common astrometric grid of pixel size 5' to yield independent pixels using the IDL routine HASTROM. \\

\section{Method}\label{method_section}

\indent The methodology used to derive the dust abundance (G/D or D/G) as a function of gas surface density is as follows. First, MW cirrus emission is estimated and removed from the FIR maps. The level of FIR emission by foreground MW cirrus can be comparable to faint emission in the periphery of the LMC or SMC, and so subtracting the foreground cirrus is particularly important for our purpose of measuring the dust abundance in very diffuse regions of the LMC and SMC. Second, the FIR dust emission is summed (or ''stacked'') over pixels in a given interval of gas surface density. A modified black body model is fit to the stacked SED to obtain the dust surface density, dust temperature, and spectral emissivity index for that gas surface density. The dust abundance (G/D) is then computed as the ratio of the gas to the dust surface density in each interval.

\subsection{Benefits of the stacking analysis on all-sky surveys}

\begin{figure*}
\includegraphics[width=8cm]{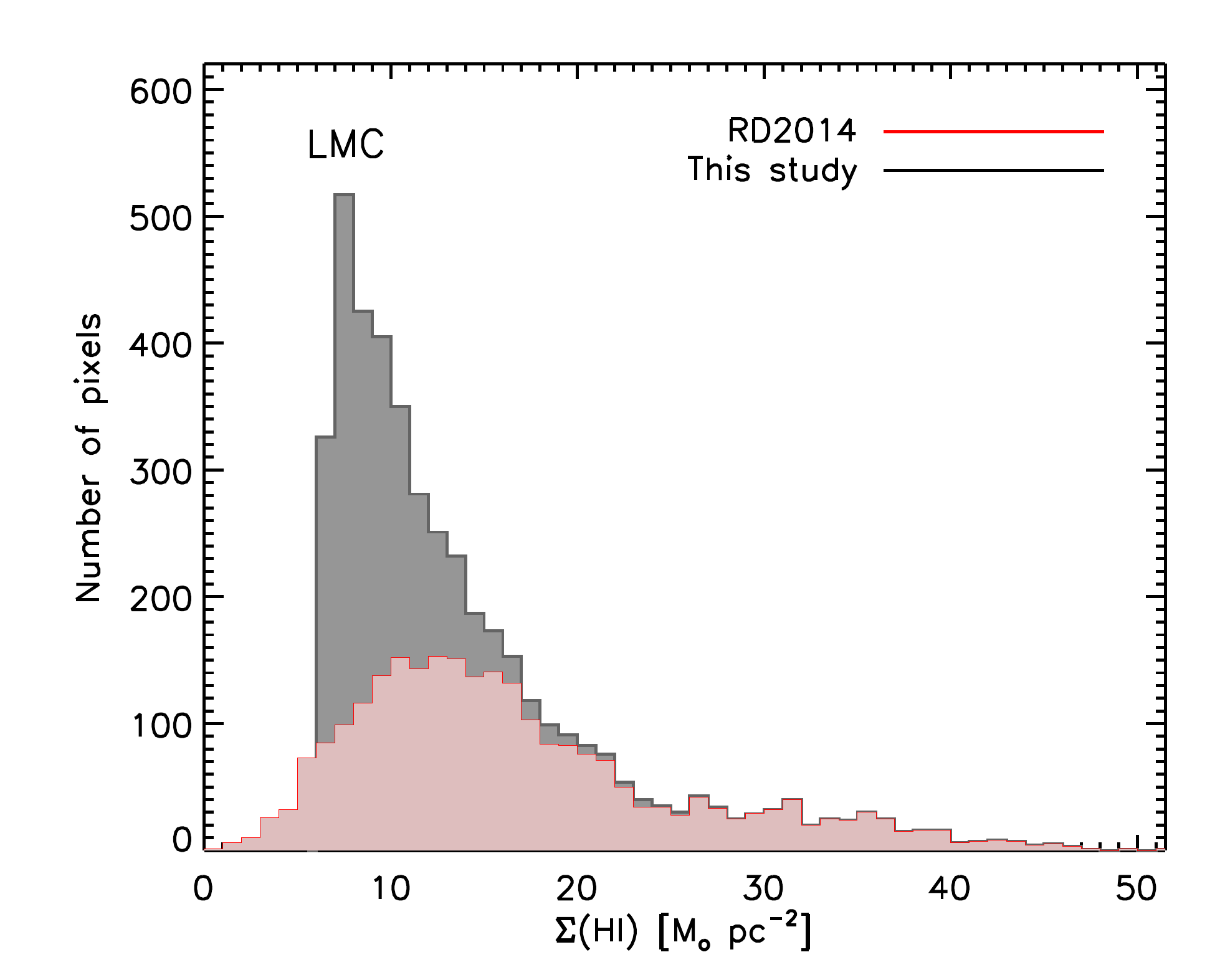}
\includegraphics[width=8cm]{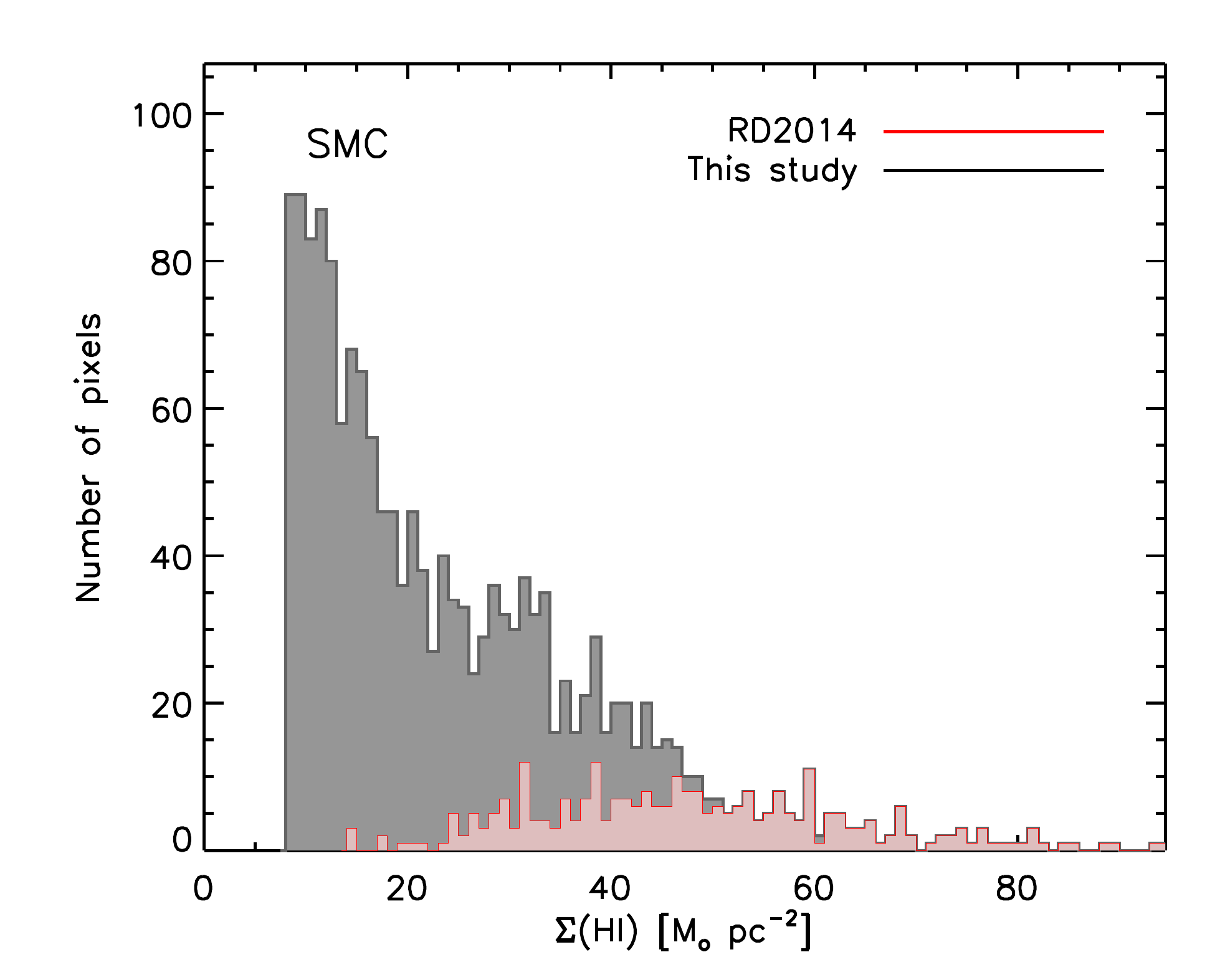}
\caption{Histogram of the \his surface density (not including Helium) of pixels included in this analysis (gray), compared to the coverage in \citet{RD2014} (red). This study includes a much greater area with low surface densities. }
\label{plot_histo_hi_coverage}
\end{figure*}

\indent In \citet{RD2014}, we used the dust surface density maps derived from Herschel data in \citet{gordon2014} convolved to 1' (15 pc) in the LMC and 98" (45 pc) in the SMC to examine the gas-to-dust ratio in the atomic and molecular ISM. \citet{RD2014} performed a linear fit to the relation between dust and gas surface densities in the regime where 1) gas is completely atomic and 2) molecular gas is traced by CO, and thus obtained a single G/D value for each regime. Figure \ref{plot_histo_hi_coverage} shows the distribution of \his surface densities in the analysis by \citet{RD2014}. The linear fits to the dust-gas relation are dominated by \his surface densities $\Sigma$(\hi) $=$ 10---20 \Msu pc$^{-2}$ ($\Sigma_g$ $=$ 14---27 \Msu pc$^{-2}$) in the LMC and  $\Sigma$(\hi) $=$ 40---60 \Msu pc$^{-2}$ ($\Sigma_g$ $=$ 55---80 \Msu pc$^{-2}$) in the SMC. The distribution of \his surface densities used in this analysis based on stacking of the Planck and IRAS data is also shown in Figure \ref{plot_histo_hi_coverage}, and extends down to much lower \his surface density in the SMC over a much larger area (as shown by the number of included pixels). Similarly in the LMC, a substantially larger area in diffuse regions is included in this new analysis.  The additional coverage is illustrated in Figure \ref{figs_3col}, which shows a 3-color image of the LMC and SMC with the area included in each analysis.  \\
\indent  In addition to improvements on the zero-point calibration (see Section \ref{introduction}), the stacking analysis performed in this paper is thus focused on the variations of G/D as a function of surface density, down to surface densities substantially lower than \citet{RD2014}, and on scales larger than in \citet{RD2014} \citep[75 pc/90 pc in the LMC/SMC here vs 15 pc/45 pc in][]{RD2014}. Ultimately, we were able to measure the variations of G/D with surface density between $\Sigma_g$ $\sim$ 8 \Msu pc$^{-2}$ and 100 \Msu pc$^{-2}$ on 75 pc/90 pc scales, instead of a bimodal estimation of G/D in the atomic and CO-bright ISM in \citet{RD2014}. \\
\indent Lastly, we have considerably improved the subtraction of foreground MW cirrus emission compared to \citet{gordon2014}. In \citet{gordon2014}, the MW cirrus subtraction was done by converting the MW \his column density toward the LMC and SMC to FIR emission assuming conversion factors modified from \citet{compiegne2011}, derived from the average diffuse MW SED. This resulted in significant over-subtraction in the outskirts of the LMC and SMC, likely due to the fact that the color of the MW cirrus varies with position. Here, we empirically derive the conversion factors specifically for each region (LMC and SMC), which allows us to reduce the systematic uncertainties in our FIR maps and accurately estimate the diffuse dust emission from the LMC and SMC.

\subsection{Milky Way Cirrus Subtraction}\label{cirrus_section}

\begin{figure*}
\centering
\includegraphics[width=8cm]{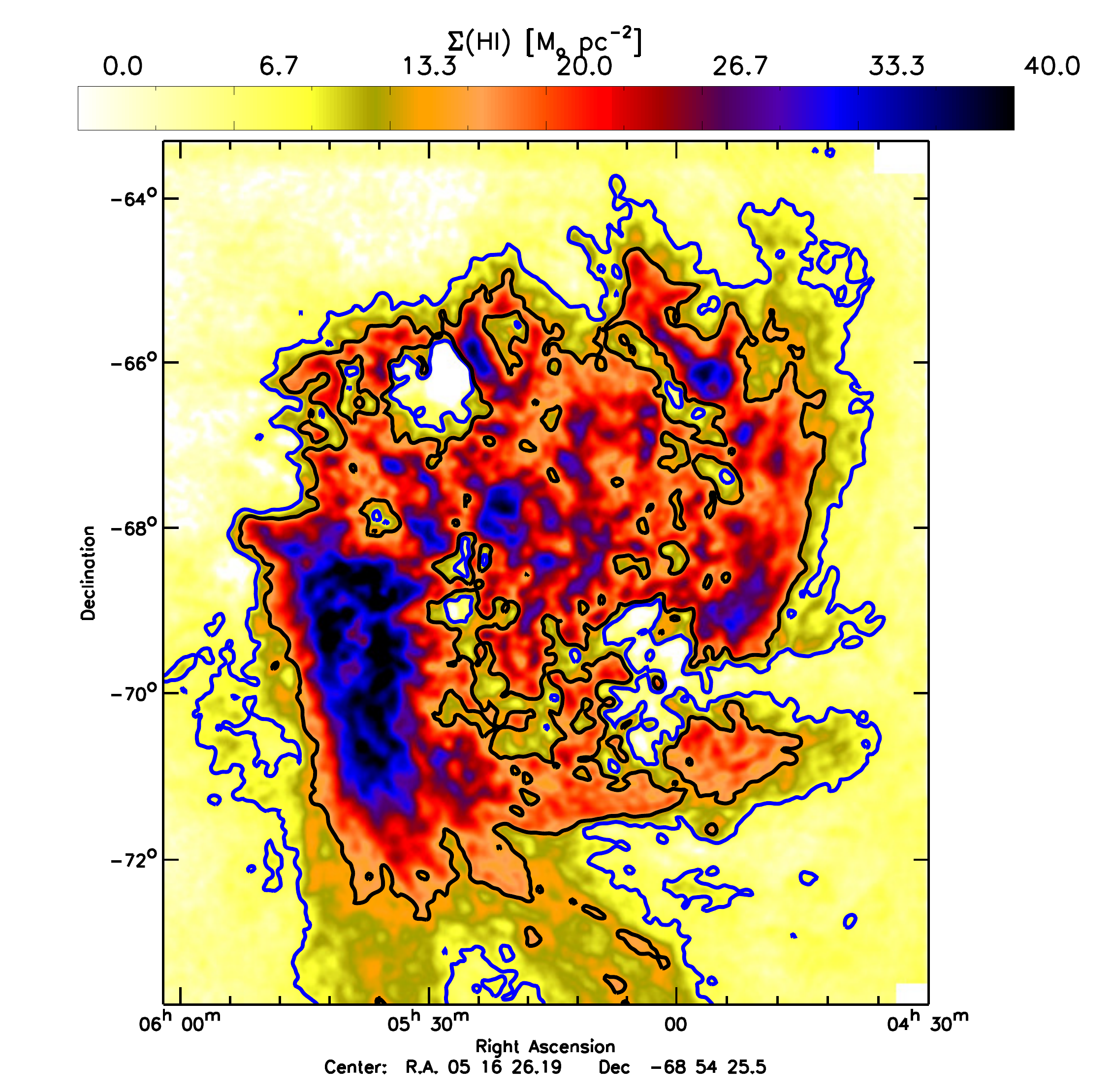}
\includegraphics[width=8cm]{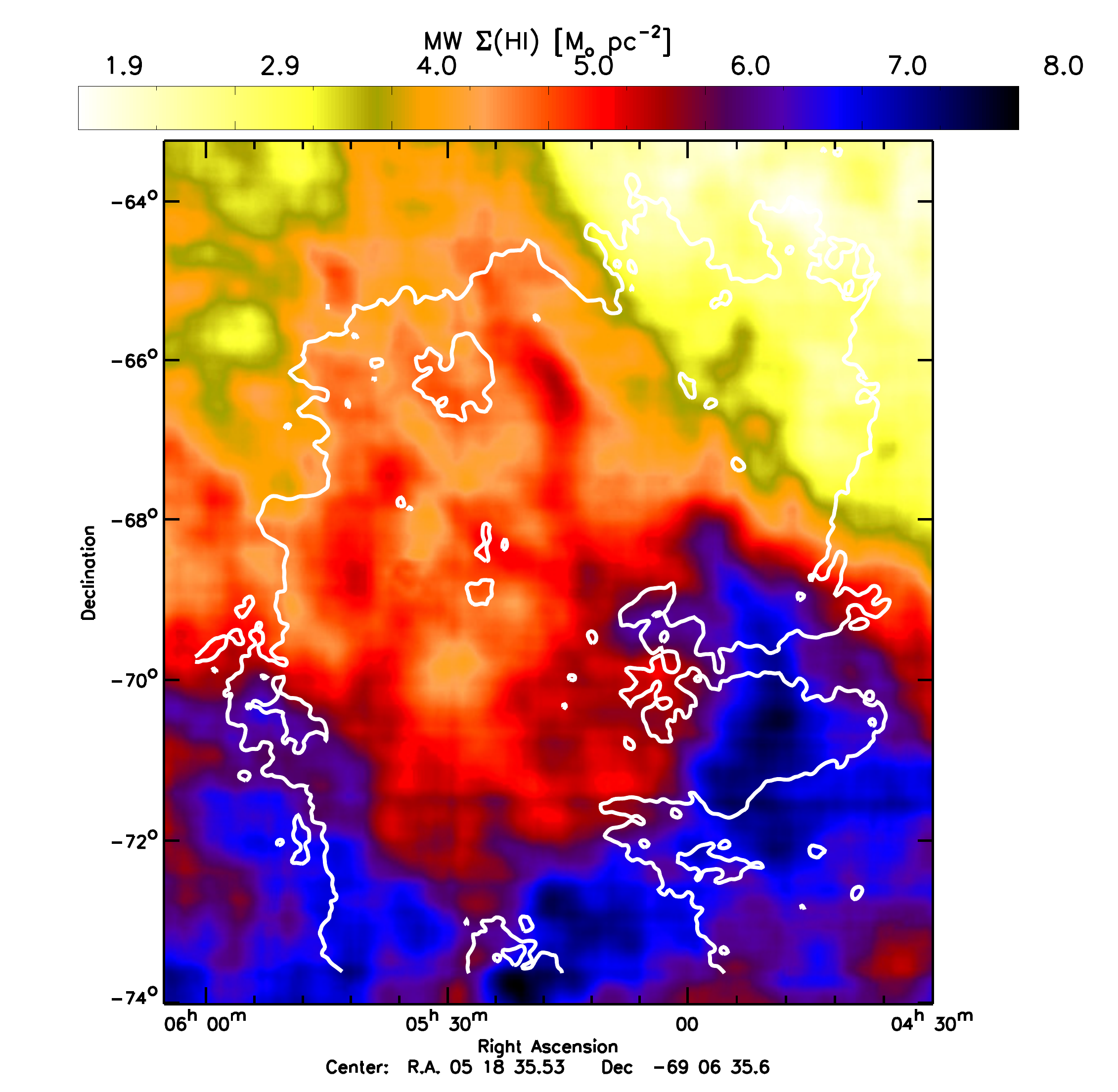}
\includegraphics[width=8cm]{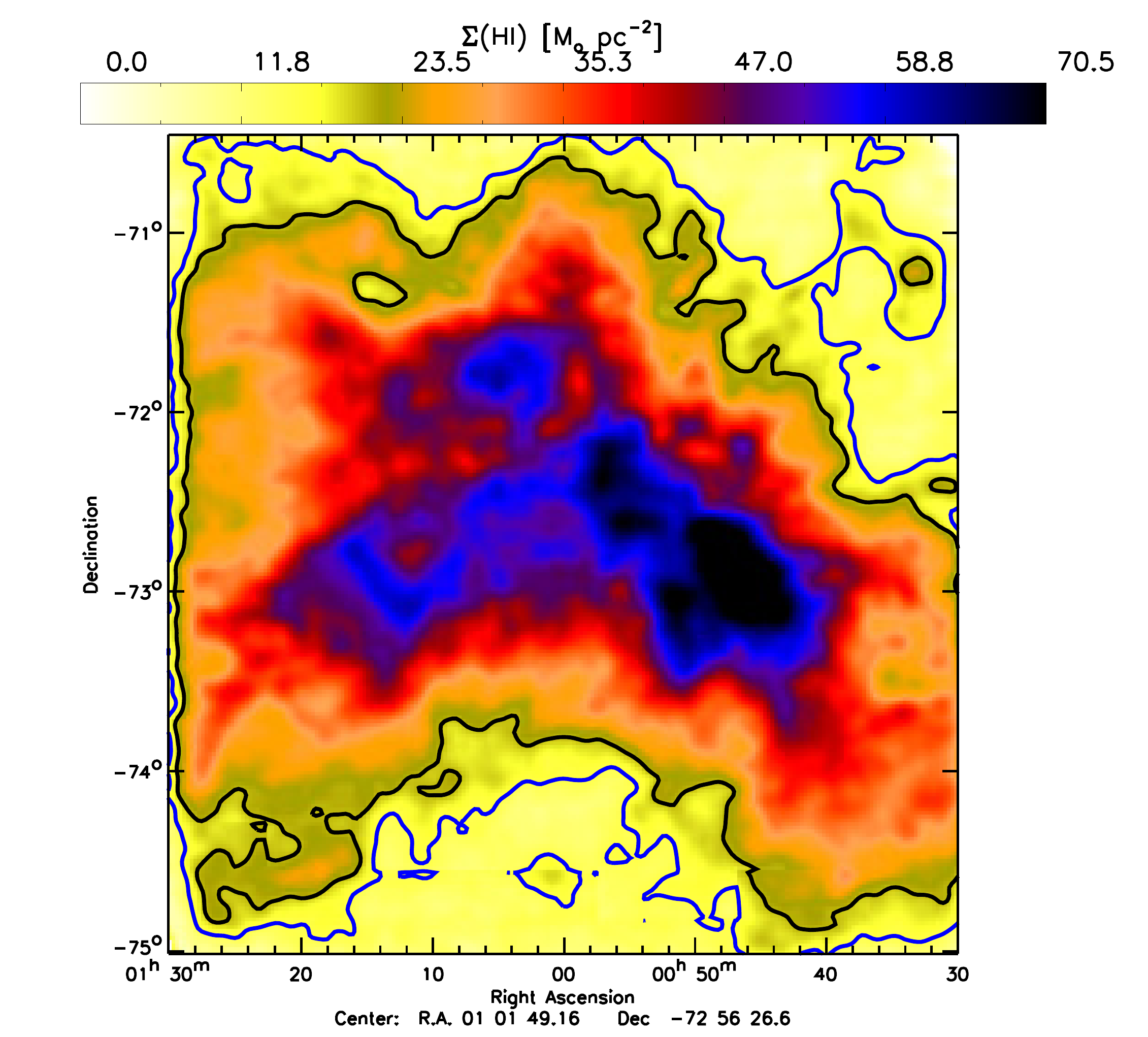}
\includegraphics[width=8cm]{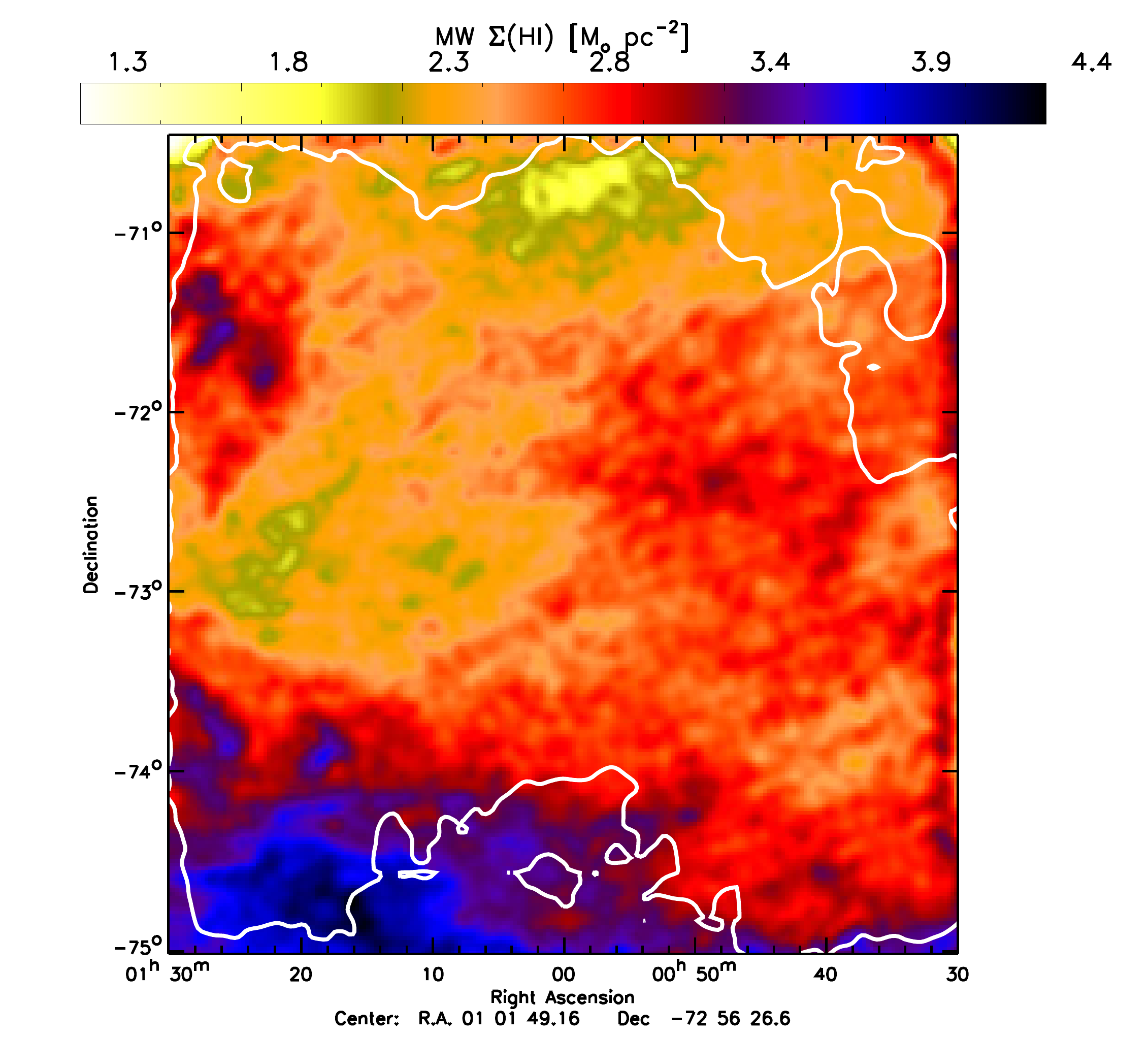}
\caption{HI surface density maps toward the LMC (top) and SMC (bottom) at LMC/SMC velocities (left) and MW velocities (right). In the left panels, the blue contours indicate the $\Sigma$(\hi) $=$ 4 \Msu pc$^{-2}$ level (not including Helium), used to define LMC/SMC regions free of detectable gas and estimate the MW cirrus emission (Section \ref{cirrus_section}). The black contours show the $\Sigma$(\hi) $=$ 6.2 \Msu pc$^{-2}$ \his surface density level (corresponding to $\Sigma_g$ $=$ 8.4 \Msu pc$^{-2}$), above which the G/D can be robustly determined (Section \ref{modeling_section} and Figure \ref{plot_gas_dust_emission}). In the right panels, the white contours are identical to the blue contours from the left panels. }
\label{hi_maps}
\end{figure*}

\begin{figure*}
\includegraphics[width=8cm]{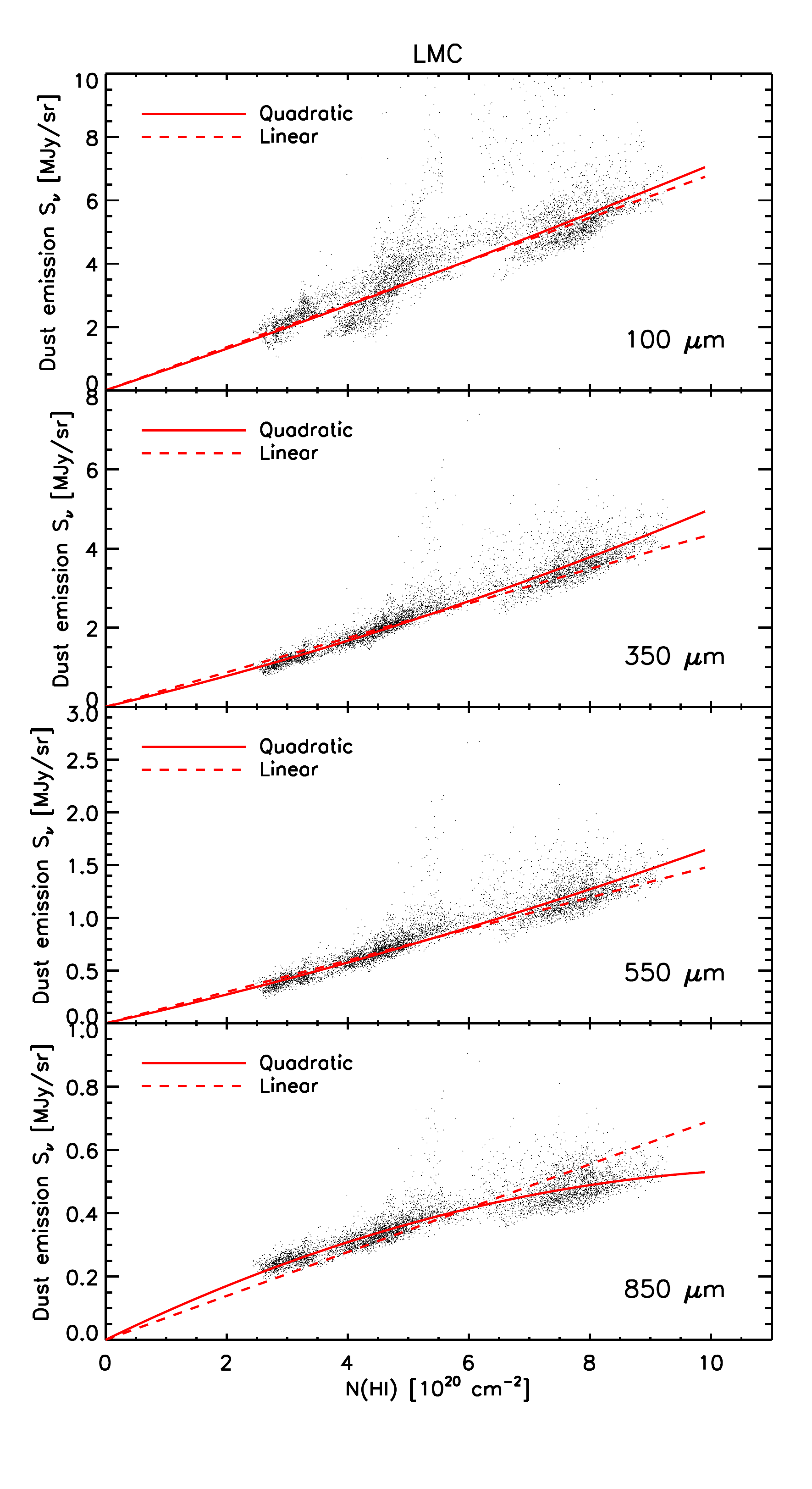}
\includegraphics[width=8cm]{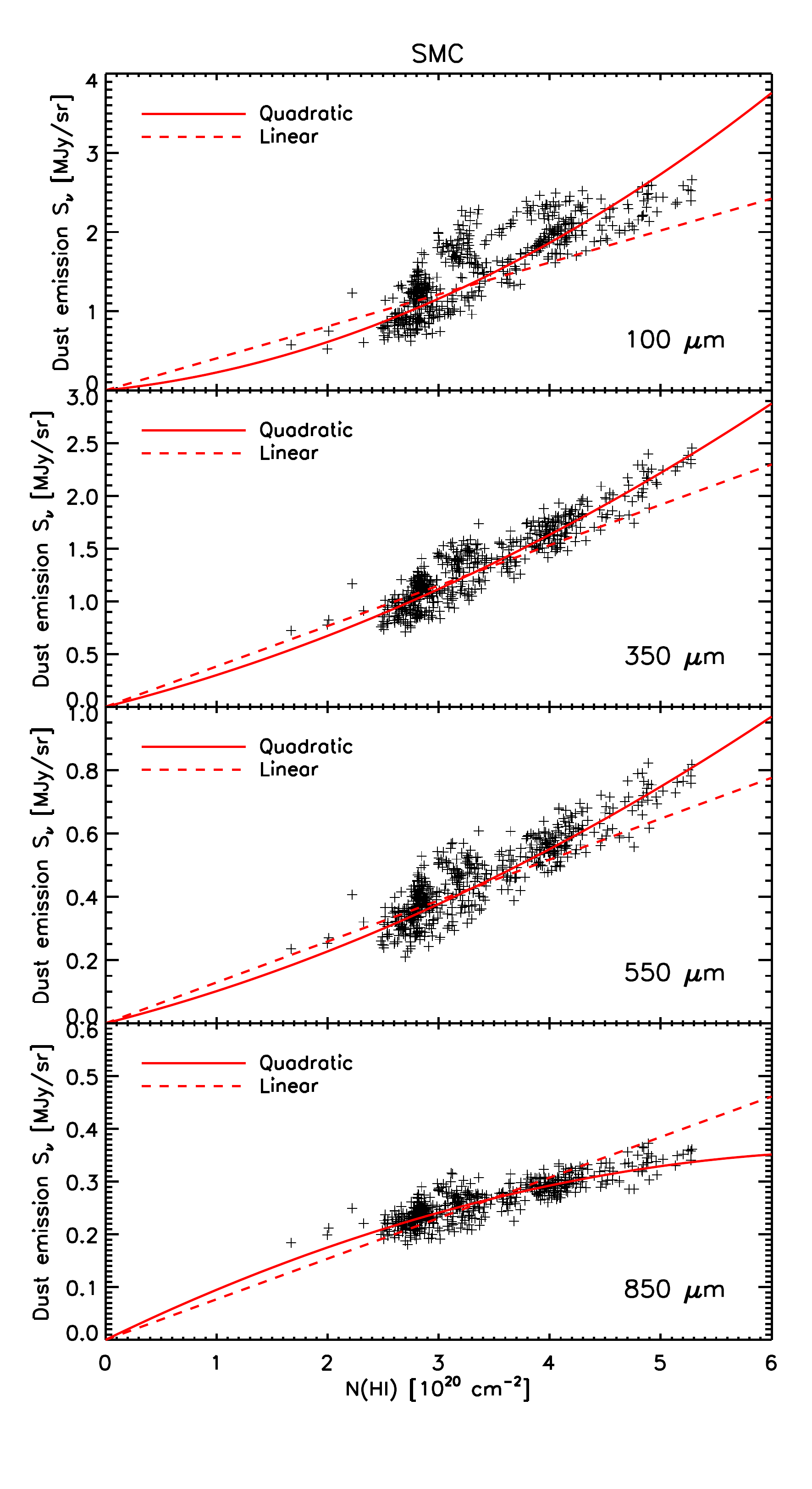}
\caption{Dust emission in the IRAS 100 $\mu$m and Planck 350 $\mu$m, 550 $\mu$m, and 850 $\mu$m as a function of MW cirrus HI surface density, toward the LMC (left) and SMC (right). Linear and quadratic fits to their relation are shown in blue  as dashed and solid lines, respectively. The coefficients of the fits are listed in Table 1 }
\label{cirrus_fits}
\end{figure*}

\indent MW dust cirrus emission occurs in the foreground of the LMC and SMC dust emission at a level of a few MJy/sr at the peak of the SED (in the range 100-350 $\mu$m). We estimate the detailed structure of the MW dust emission in the IRAS and Planck bands using the surface density map of \his in the MW, $\Sigma_{\mathrm{MW}}$(\hi), in the direction of the LMC \citep{staveleysmith03} and SMC \citep{stanimirovic1999}. The MW \his surface density was obtained by integrating the \his cubes over MW velocities. The cirrus dust emission, $S_{\nu}^{\mathrm{MW}}$ in MJy/sr, in each band was obtained by applying either a linear ($S_{\nu}^{\mathrm{MW}}=a_1\Sigma_{\mathrm{MW}}$(\hi)) or quadratic ($S_{\nu}^{\mathrm{MW}}=a_1\Sigma$(\hi) $+$ $a_2\Sigma_{\mathrm{MW}}$(\hi)$^2$)  function to the MW \his surface density maps, $\Sigma_{\mathrm{MW}}$(\hi). We constrained the linear and quadratic coefficients ($a_1$ and $a_2$) by examining the relation between $S_{\nu}^{\mathrm{MW}}$ and $\Sigma_{\mathrm{MW}}$(\hi) in ''cirrus calibration regions", where no \his emission at LMC or SMC velocities is detected to within the sensitivity limits of our maps, corresponding to pixels with \his surface densities below 4 \Msu pc$^{-2}$. The contours corresponding to this threshold are shown in Figure \ref{hi_maps}.  In those regions (outside the contours), the paucity of gas originating in the LMC/SMC guarantees the absence of dust emission from those galaxies, particularly given their low metallicty. Therefore, dust emission in those regions originates solely from Galactic foreground, and $S_{\nu}$ $=$ $S_{\nu}^{\mathrm{MW}}$, where $S_{\nu}$ is the emission in the maps. Figure \ref{cirrus_fits} shows the relation between MW \his surface density and cirrus dust emission in each band and toward each galaxy, as well as the linear and quadratic fits to those relations. The coefficients of the fits are given in Table 1. \\
\indent We note that the cirrus emission exhibits noticeable differences in color between the sight-lines toward the LMC and SMC, particularly in colors involving the 100 $\mu$m band. Hence, an attempt at fitting the relation between MW \his and dust emission simultaneously for both sight-lines failed, because it resulted in an over-subtraction in the IRAS 100 $\mu$m band in the SMC. Hence, we estimate the cirrus emission in the LMC and SMC sight-lines separately. \\
\indent Figure \ref{cirrus_fits} shows that a quadratic function provides a better description of the relation between MW \his surface density and cirrus dust emission. In the following, we therefore use the maps obtained from MW cirrus subtraction based on  a quadratic fit as our fiducial case. This result is not unexpected.  As the gas surface density increases, the temperature and spectral emissivity index of the dust vary, resulting in a non-linear relation between gas surface density and FIR emission. \\

\begin{deluxetable*}{ccccccc}
\tabletypesize{\scriptsize}
\tablecolumns{7}
\tablewidth{\textwidth}
\tablecaption{Coefficients of the fits to the relation between MW \his column density and cirrus dust emission}
\tablenum{1}
 
 \tablehead{
 & \multicolumn{3}{c}{LMC} & \multicolumn{3}{c}{SMC}}

 \startdata
 & Linear & \multicolumn{2}{c}{Quadratic}  & Linear & \multicolumn{2}{c}{Quadratic} \\

Band & $a_1$ & $a_1$ & $a_2$& $a_1$ & $a_1$ & $a_2$ \\

&&& &&&\\
\hline
&&&&&&\\



100 & 0.681$\pm$8.9$\times10^{-4}$ & 0.642$\pm$2.8$\times10^{-3}$ & 0.00706$\pm$4.7$\times10^{-4}$ & 0.404$\pm$1.8$\times10^{-3}$ & 0.145$\pm$9.0$\times10^{-3}$ & 0.0802$\pm$2.7$\times10^{-3}$ \\
350 & 0.436$\pm$5.5$\times10^{-4}$ & 0.362$\pm$1.6$\times10^{-3}$ & 0.0138$\pm$2.9$\times10^{-4}$ & 0.383$\pm$1.7$\times10^{-3}$ & 0.265$\pm$8.6$\times10^{-3}$ & 0.0357$\pm$2.6$\times10^{-3} $\\
550 & 0.149$\pm$1.9$\times10^{-4}$ & 0.129$\pm$5.6$\times10^{-4}$ & 0.00370$\pm$9.9$\times10^{-5}$ & 0.129$\pm$5.7$\times10^{-4}$ & 0.0897$\pm$2.9$\times10^{-3}$ & 0.0120$\pm$8.6$\times10^{-4}$ \\
850 & 0.0693$\pm$8.8$\times10^{-5}$ & 0.0932$\pm$2.9$\times10^{-4}$ & -0.00400$\pm$4.7$\times10^{-5}$ & 0.0769$\pm$3.4$\times10^{-4}$ & 0.102$\pm$1.7$\times10^{-3}$ & -0.00726$\pm$4.8$\times10^{-4}$\\

   \enddata
  \label{cirrus_fits_coefs}
  \tablecomments{The linear and quadratic relations are described by $S_{\nu}^{\mathrm{MW}}=a_1 N_{20}^{\mathrm{MW}}$(\hi) and $S_{\nu}^{\mathrm{MW}}=a_1 N_{20}^{\mathrm{MW}}$(\hi) $+$ $a_2 N_{20}^{\mathrm{MW}}$(\hi)$^2$, respectively, where $N_{20}^{\mathrm{MW}}$ is the MW \his column density in units of $10^{20}$ H cm$^{-2}$. The $a$ coefficients assume the emission is given in MJy/sr, and thus $a_1$ has units MJy sr (10$^{20}$ H cm$^{-2}$)$^{-1}$ and $a_2$ has units MJy sr (10$^{20}$ H cm$^{-2}$)$^{-2}$  }

\end{deluxetable*}

\subsection{Stacking of the dust emission}

\begin{figure}
\includegraphics[width=8cm]{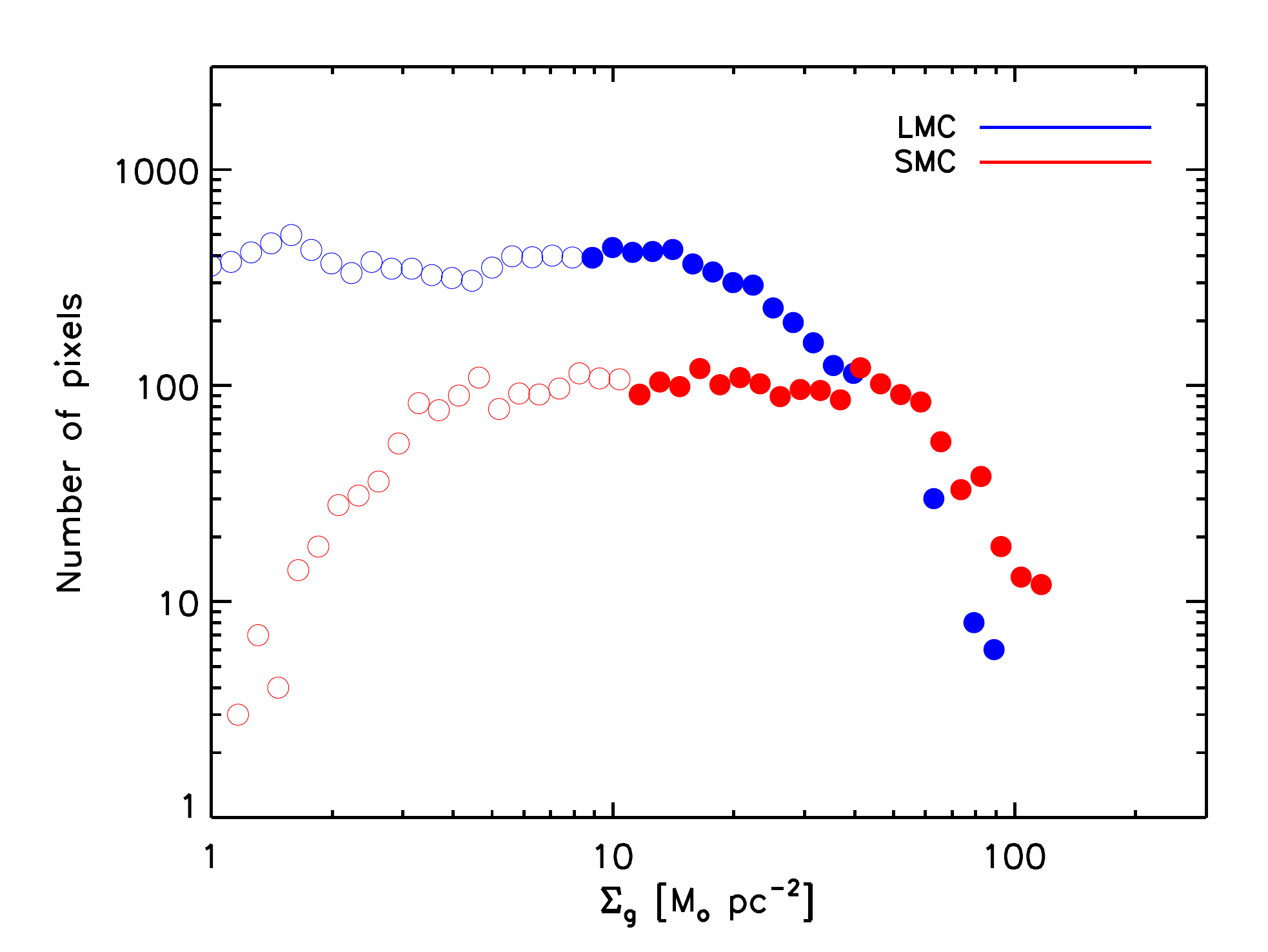}
\caption{Number of pixels in each bin of gas surface density in the LMC (blue) and SMC (red). The open circles correspond to bins that are not used in this analysis because the stacked dust emission cannot be reliably estimate (Section \ref{modeling_section}), while filled circles correspond to bins used to estimate the G/D and other dust parameters. }
\label{plot_counts}
\end{figure}

\begin{figure*}
\includegraphics[width=\textwidth]{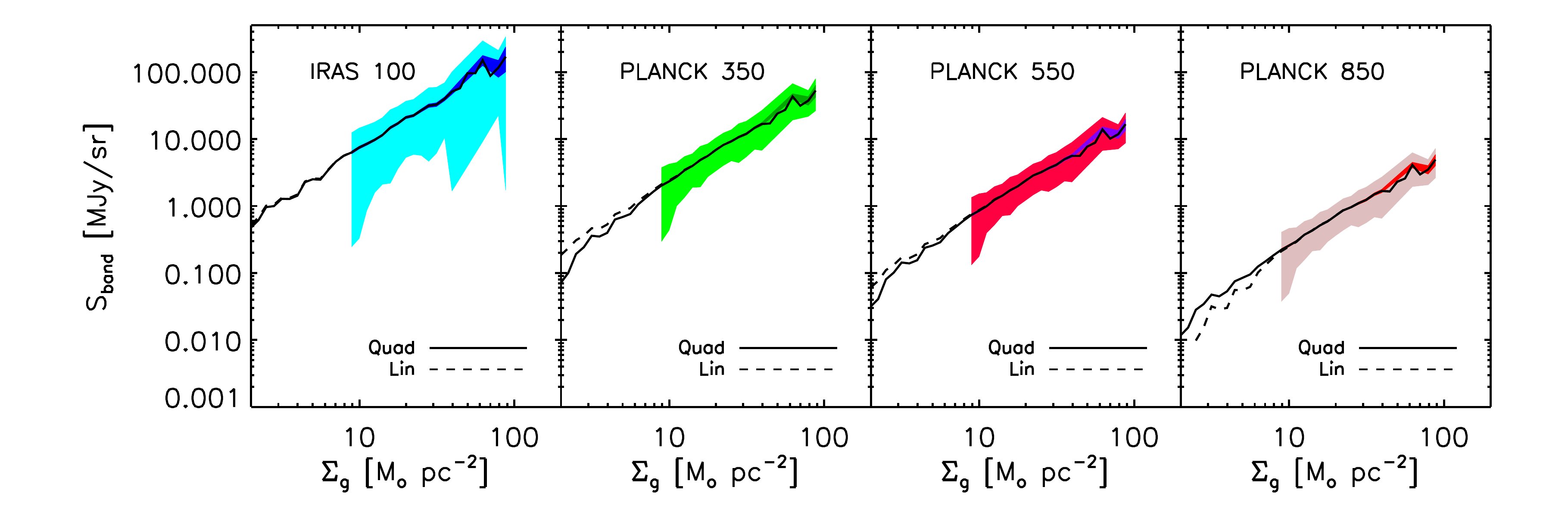}
\includegraphics[width=\textwidth]{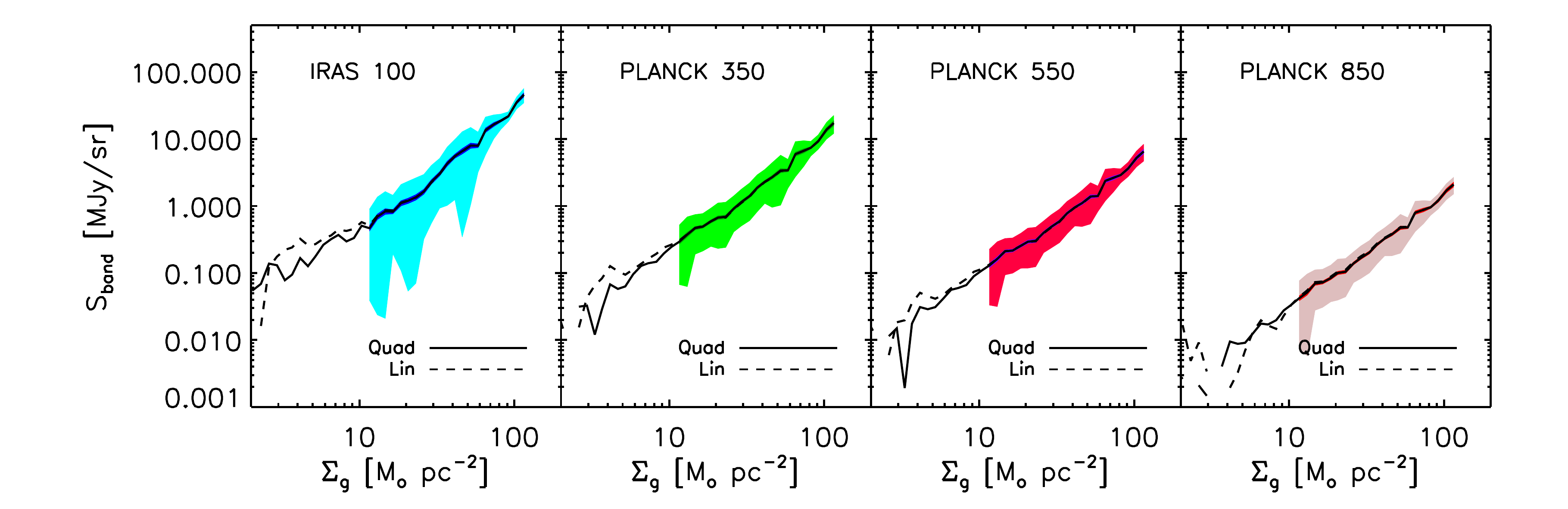}

\caption{Dust emission from the LMC (top) and SMC (bottom) in the IRAS 100 $\mu$m and Planck 350 $\mu$m, 550 $\mu$m, and 850 $\mu$m as a function of  gas surface density.  The outer (lighter) band shows the standard deviation in the stacked dust emission, while the inner (darker) band shows the error on the mean. Note that the inner darker band is narrow and hard to separate from the solid line. The standard deviation and error on the mean are only shown for bins used in the analysis, i.e., bins where the standard deviation of the emission in each bin is greater than the mean emission in that bin. The solid and dashed lines correspond to dust emission after subtracting quadratic and linear fits to the MW foreground cirrus emission, respectively. }
\label{plot_gas_dust_emission}
\end{figure*}

\indent Once the foreground MW emission is removed, we proceed with the stacking analysis of the dust emission. The stacking analysis allows us to beat the noise in diffuse regions with faint FIR emission and measure dust emission at surface densities much lower than previously observed in \citet{gordon2014} and \citet{RD2014}. Since dust and gas are mixed in the ISM, it is expected that the dust emission correlates best with gas surface density. We therefore stack the dust emission in bins of gas surface density. \\
\indent The pixels of the dust maps are separated into bins of gas surface density. Each bin of gas surface density is 0.05 dex wide. The number of pixels in each bin, shown in Figure \ref{plot_counts}, varies between 5 and 2000. The decreasing number of pixels in bins of increasing surface density is due to the hierarchical (fractal) nature of the ISM. The diffuse, low surface density ISM occupies a much bigger volume that its dense counterpart. In order to sample the high range of the surface density, one has to use bins of equal surface density width, rather than the statistically advantageous equal number of pixels in each bin. \\
\indent For each gas surface density bin, we compute the mean, standard deviation, and error on the mean of the dust emission in each band, $S_{\mathrm{band}}^{\mathrm{mean}}$, $S_{\mathrm{band}}^{\mathrm{stdv}}$, and $S_{\mathrm{band}}^{\mathrm{err}}$ $=$ $S_{\mathrm{band}}^{\mathrm{stdv}}$/$\sqrt{N_{\mathrm{stack}}}$, from all the $N_{\mathrm{stack}}$ pixels in each gas surface density interval. We have verified that the mean gas surface density in each bin equals the center of the surface density bin to within 5\%. The stacked dust emission in each band is plotted as a function of gas surface density in Figure \ref{plot_gas_dust_emission}.\\

\subsection{Modeling of the dust emission}\label{modeling_section}

\indent We then fit a modified black body model \citep[SMBB model in ][]{gordon2014} to the stacked dust emission SED, $S_{\mathrm{band}}^{\mathrm{mean}}$, at 100, 350, 550, 850 $\mu$m in each bin of gas surface density. The modified black body function takes the form:

\begin{equation}
S_{\nu}(\lambda) = (2.0891\times10^4)\kappa_{0} \left (\frac{\lambda_0}{\lambda} \right )^{\beta}\Sigma_d B_{\nu}(\lambda, T_d)
\label{smbb_model}
\end{equation}

\noindent where  $S_{\nu}$ is the surface brightness, $B_{\nu}(\lambda, T_d)$ is the Planck function (and  $S_{\nu}$ and $B_{\nu}(\lambda, T_d)$  have the same dimensions and units), $\kappa_0$ is the dust emissivity of the big grains at wavelength $\lambda_0$ in cm$^2$ g$^{-1}$, and $\Sigma_d$ is the dust surface density in \Msu pc$^{-2}$. $\Sigma_d$, $T_d$, and $\beta$ are left as free parameters. We note that we also performed the dust SED modeling with the model of a black body modified by a broken-emissivity law \citep[BEMBB model described by Equations 7, 9, and 10 in ][]{gordon2014}. While the dust parameters we derived (surface density, abundance, temperature, spectral emissivity index) are consistent between the two modeling approaches, uncertainties on the derived parameters were significantly larger for the BEMBB model due to the numbers of bands (4 in this study) being lower than the number of parameters (5 for the BEMBB model). Given the lack of scientific benefit in using the BEMBB model (same results, larger error bars) and, as we will see,  the absence of significant residuals with the SMBB model, we only use the SMBB model described in Equation \ref{smbb_model} in the rest of this paper. \\
\indent We calibrate $\kappa_0$ at $\lambda_0$ $=$ 160 $\mu$m using the Milky Way diffuse dust SED from \citet{compiegne2011} and the knowledge of the dust-to-gas ratio toward this sight-lines from depletions \citep{jenkins09}. The self-calibration of $\kappa_0$ is explained in details in Section \ref{kappa_cal_section}, and leads to $\kappa_0$ $=$ 12.4 cm$^{2}$ g$^{-1}$ at $\lambda_0$ $=$ 160 $\mu$m, consistent with dust models by \citet{draine2007, draine2014, zubko2004}.\\
\indent The model is integrated with the response functions for the IRAS 100 $\mu$m and Planck 350 $\mu$m, 550 $\mu$m, and 850 $\mu$m bands, $R_E(\lambda)$. Since the IRAS and Planck missions assume a $\nu_0$/$\nu$ spectral energy distribution to compute the band flux, the models are computed using the following (with $\nu$ = $c/\lambda$):

\begin{equation}
S_{\mathrm{band}}^{\mathrm{mod}} = \frac{\int R_E(\nu) S_{\nu}(\nu) d\nu }{\int R_E(\nu) \left ( \frac{\nu_{\mathrm{eff}}}{\nu} \right) d\nu}
\end{equation}

\noindent where $\nu_{\mathrm{eff}}$ is the effective frequency of the band. The corresponding wavelengths are 95.30 $\mu$m, 347.8 $\mu$m, 538.1 $\mu$m, and 830.4 $\mu$m for the IRAS 100 $\mu$m and Planck 350 $\mu$m, 550 $\mu$m, and 850 $\mu$m bands, respectively.\\
\indent We pre-computed a grid of model SEDs in the IRAS and Planck bands ($S_{\mathrm{band}}^{\mathrm{mod}}$) with dust parameters $\Sigma_d$ $=$ $10^{-6}$ --- 1 \Msu pc$^{-2}$ in steps of 0.02 dex, $T_d$ $=$ 5---49 K in steps of 0.5 K, and $\beta$ $=$ 0---3 in steps of 0.1. For each gas surface density bin, we fit the model grid to the observed, stacked SED ($S_{\mathrm{band}}^{\mathrm{mean}}$) using a probabilistic approach as in the DustBFF dust model code \citep{gordon2014}. The dust fitting algorithm uses the base 10 logarithm of the dust surface density, since its PDF has been shown to be approximately gaussian in the log \citep{gordon2014}, while the PDF of the dust surface density exhibits substantial wings, which could bias the estimate of the best-fit dust surface density. In other words, we compute the PDF and expectation value of $\log_{10}(\Sigma_d$). We compute the probability of each model as:

\begin{equation}
P(\logten \Sigma_d, T_d, \beta|S_{\mathrm{band}}^{\mathrm{mean}}) = \frac{1}{Q} \exp{ \left ( -\frac{1}{2} \chi^2 \right )} 
\end{equation}

\noindent where

\begin{equation}
\chi^2 = \left ( S_{\mathrm{band}}^{\mathrm{mod}} -  S_{\mathrm{band}}^{\mathrm{mean}}  \right )  \mathbb{C}^{-1}  \left ( S_{\mathrm{band}}^{\mathrm{mod}} -  S_{\mathrm{band}}^{\mathrm{mean}}  \right ) 
\end{equation}

\begin{equation}
Q = (2\pi)^n \mathrm{det} |\mathbb{C}|
\end{equation}

\noindent $n$ is the number of bands (4 in our case). $\mathbb{C}$ is the covariance matrix and is the sum of the covariance matrices corresponding to calibration, background, and measurement errors, respectively: $\mathbb{C}$ $=$ $\mathbb{C}^{\mathrm{cal}}$ $+$ $\mathbb{C}^{\mathrm{bkg}}$ $+$ $\mathbb{C}^{\mathrm{meas}}$. the covariance matrix is calculated for each SED in each bin of gas surface density. The covariance matrix corresponding to measurement errors is diagonal (no covariance between bands) and the diagonal terms are given by the error on the mean dust emission in each band for a given bin of surface density: $\mathbb{C}^{\mathrm{meas}}_{ii}$ $=$ $S_i^{\mathrm{err}}$. The covariance matrix corresponding to residual uncertainties on the background reflects variations in the CIB and residual emission from the MW cirrus emission (the cirrus subtraction is not perfect). $\mathbb{C}^{\mathrm{bkg}}$ is estimated in regions with no detectable dust emission from the LMC and SMC. We make the same assumption as for the MW cirrus estimation in Section \ref{cirrus_section}, and use regions where the \his emission from the LMC and SMC corresponds to \his surface densities below 4 \Msu pc$^{-2}$.   $\mathbb{C}^{\mathrm{bkg}}$ is given by:

\begin{equation}
\mathbb{C}^{\mathrm{bkg}}_{ij} = \frac{\sum_{k=0} ^{N_r  -1}   \left (  S_i^k - <S_i> \right ) \left (S_j^k - <S_j> \right )   }{N_{\mathrm{stack}} N_r}
\end{equation}

\noindent where $S_i^k$ is the observed surface brightness in band $i$ and pixel $k$, $N_r$ is the number of pixels used to estimate the background fluctuations in regions where there is no detectable dust emission ($\Sigma$(\hi) $<$ 4 \Msu pc$^{-2}$), $<S_i>$ is the mean observed surface brightness in band $i$ in those $N_r$ pixels, and $N_{\mathrm{stack}}$ is the number of pixels included in the stacking analysis for a given bin of gas surface density. Thus, the background covariance matrix only depends on the gas surface density bin through $N_{\mathrm{stack}}$. The other terms are constant and estimated directly in the images. In the LMC, with a quadratic fit to the cirrus emission and subtraction, $\mathbb{C}^{\mathrm{bkg}}$ is given by:

\begin{equation}
{\footnotesize
\bf{\mathbb{C}^{\mathrm{bkg}}} = \frac{1}{N_{\mathrm{stack}}} \left [ \begin{array}{cccc}
       
       2.51  &    0.38  &   0.14  &  0.040  \\
     0.38   &   0.12 &    0.046   &   0.013  \\
     0.14  &   0.046 &    0.018 &   0.0053 \\
         0.040 &   0.013 &   0.0076 &    0.0018 \\
\end{array} \right]
}
\end{equation}

\noindent In the SMC, 

\begin{equation}
{\footnotesize
\bf{\mathbb{C}^{\mathrm{bkg}}} = \frac{1}{N_{\mathrm{stack}}} \left [ \begin{array}{cccc}
       
     0.088  &  0.035  &   0.013  &  0.0035 \\
    0.035    & 0.020   & 0.0080  &  0.0025\\
    0.013   & 0.0080   & 0.0035   & 0.0011\\
   0.0035  &  0.0025 &  0.0011 &  0.00042\\
   
\end{array} \right]
}
\end{equation}

The covariance matrix corresponding to calibration uncertainties is given by $\mathbb{C}^{\mathrm{cal}}_{ij}$ $=$ $S_i^{\mathrm{mean}}S_j^{\mathrm{mean}}$ $\bf{A}_{ij}$ and $\bf{A}_{ij}$  assumes 10\%, 6.4\%, 6.1\%, and  0.78\% absolute calibration uncertainties for the IRAS 100 $\mu$m, PLANCK 350 $\mu$m, and PLANCK 550 $\mu$m bands respectively \citep[diagonal terms, see][]{r3planck2015}. For IRAS 100 $\mu$m band, there is also a 2\% repeatability term added in quadrature with the absolute calibration uncertainty. The correlated calibration uncertainties (non-diagonal terms) are 2\% between PLANCK 350 $\mu$m and PLANCK 550 $\mu$m and 0 for the other bands. Thus, 

\begin{equation}
{\footnotesize
\bf{A} = \left [ \begin{array}{cccc}
1.036 \times 10^{-2}   &   0  &    0   &    0 \\
   0          &  4.096 \times 10^{-3} & 4.000 \times 10^{-4}   &    0 \\
       0  & 4.000 \times 10^{-4}  &   3.721 \times 10^{-3}  &     0\\
       0   &    0    &   0  &  6.0834\times 10^{-5}\\
\end{array} \right]
}
\end{equation}

\indent From the probabilities of each model, we compute the marginalized (1D) probability distribution functions (PDFs) for each dust parameters by integrating $P(\Sigma_d, T_d, \beta|S_{\mathrm{band}}^{\mathrm{mean}})$ over the other two parameters, for instance:

\begin{equation}
P(\logten \Sigma_d|S_{\mathrm{band}}^{\mathrm{mean}}) = \int_{T_d, \beta} P(\logten \Sigma_d, T_d, \beta|S_{\mathrm{band}}^{\mathrm{mean}})
\end{equation}

\noindent The best dust parameters $\logten \Sigma_d, T_d, \beta$ correspond to the expectation value of the corresponding 1D PDF. The error the parameter estimation corresponds to the square root of the variance of the 1D PDF. For instance, for the dust surface density:

\begin{equation}
\left ( \logten \Sigma_d \right )^{\mathrm{best}} = \int_{\Sigma_d} P(\logten \Sigma_d|S_{\mathrm{band}}^{\mathrm{mean}})  \Sigma_d d \logten \Sigma_d
\end{equation}

\begin{multline}
\sigma \left ( \logten \Sigma_d \right )^2 = \int_{\Sigma_d} P(\logten \Sigma_d|S_{\mathrm{band}}^{\mathrm{mean}})  \times \\
\left (\logten \Sigma_d  - \left ( \logten \Sigma_d \right )^{\mathrm{best}} \right )^2 d \logten \Sigma_d
\end{multline}

\begin{figure}
\includegraphics[width=8cm]{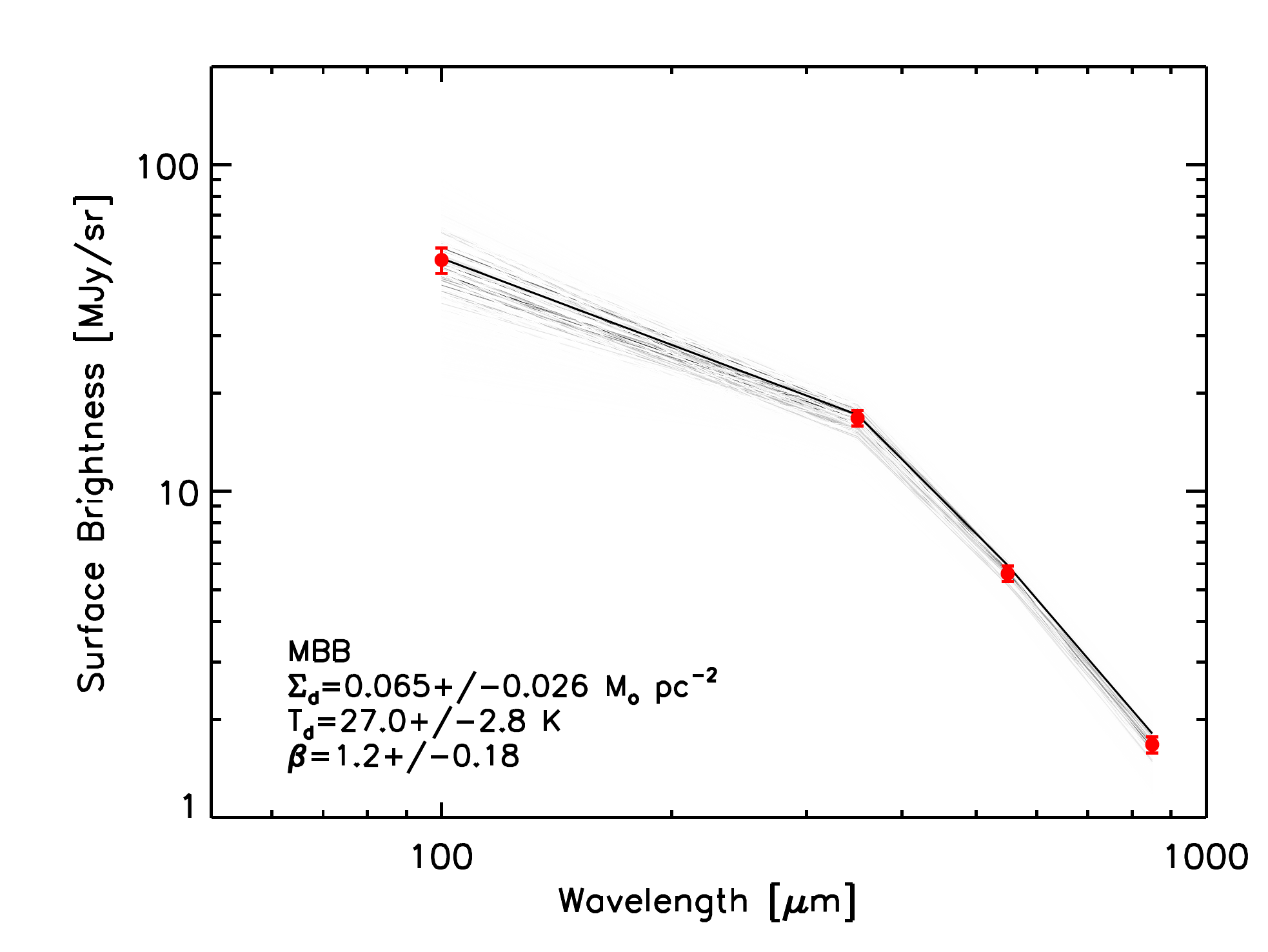}
\includegraphics[width=8cm]{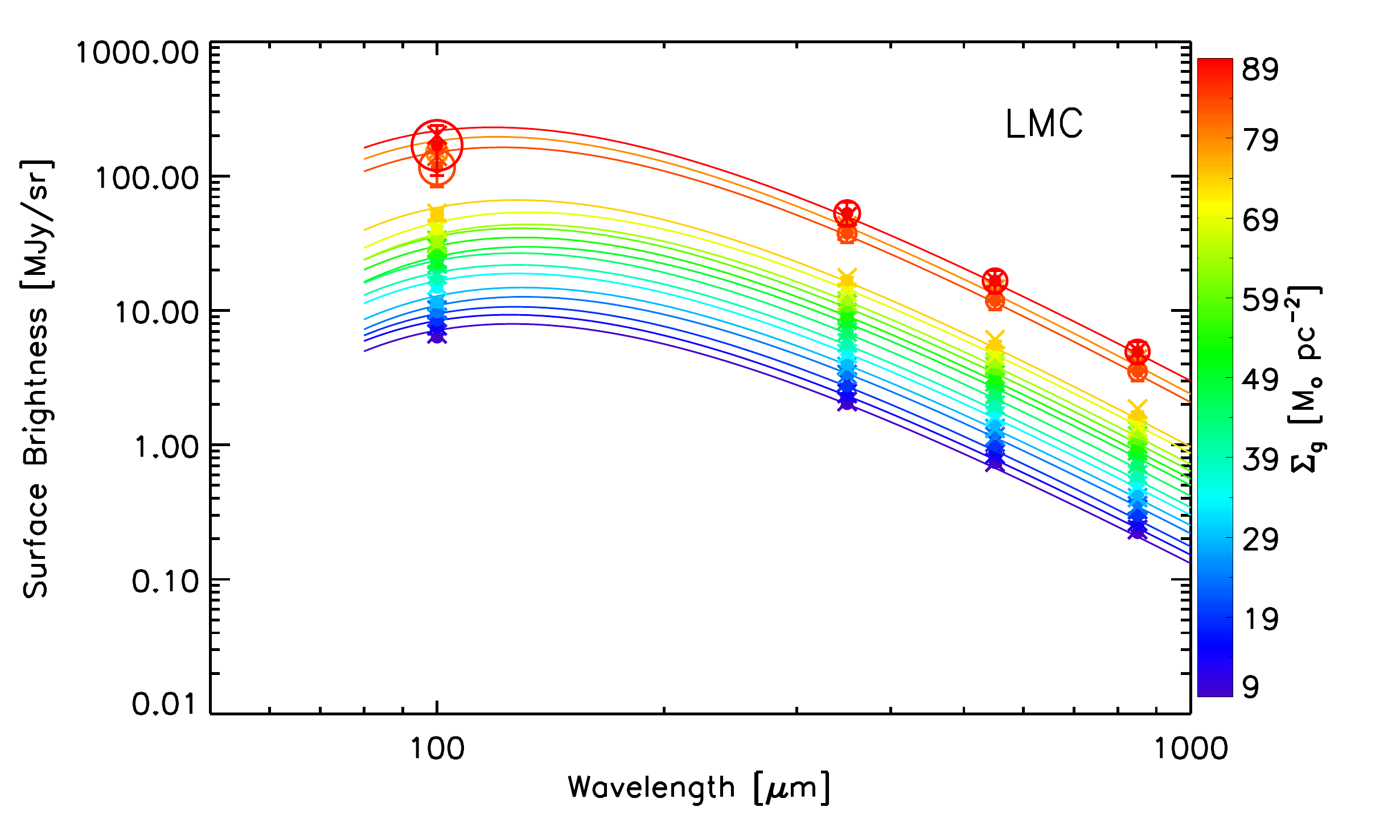}
\includegraphics[width=8cm]{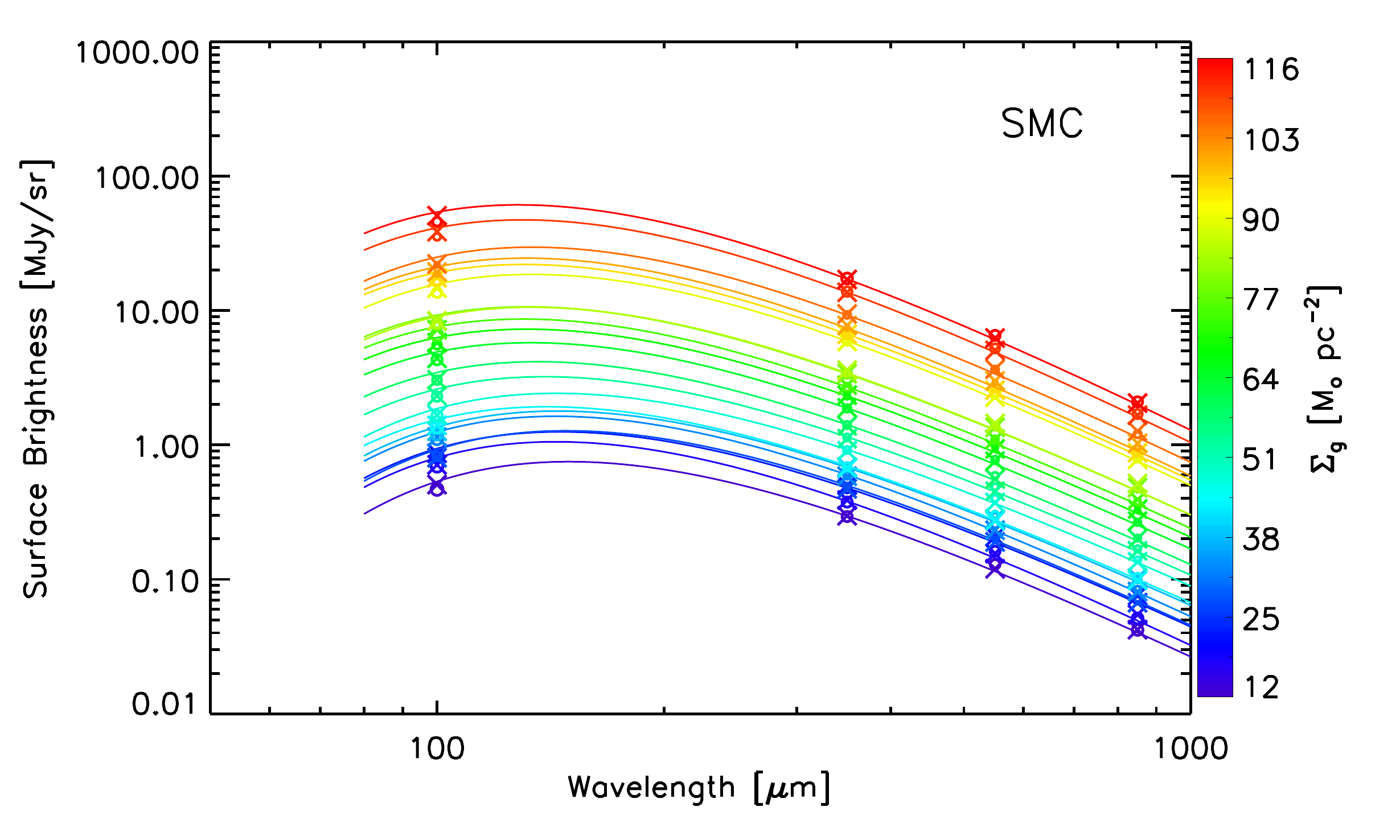}

\caption{(Top) Example of stacked dust SED for $\Sigma_g$ $=$ 46 \Msu pc$^{-2}$. The red points correspond to the stacked SED (the data). The gray lines show models randomly drawn from the marginalized 1D PDFs. The darkness of the line is proportional to the probability of the model to fit the observed SED. The thick black line corresponds to the best model, with parameters equal to the expectation value of the 1D PDFs. (Middle) SEDs (open circles, radius equal to uncertainty), best-fit models (crosses), and corresponding monochromatic models (not including response functions, lines) for all the gas surface density bins in the LMC. (Bottom) Same as above but for the SMC. }
\label{plot_sed}
\end{figure}

\begin{figure}
\includegraphics[width=8cm]{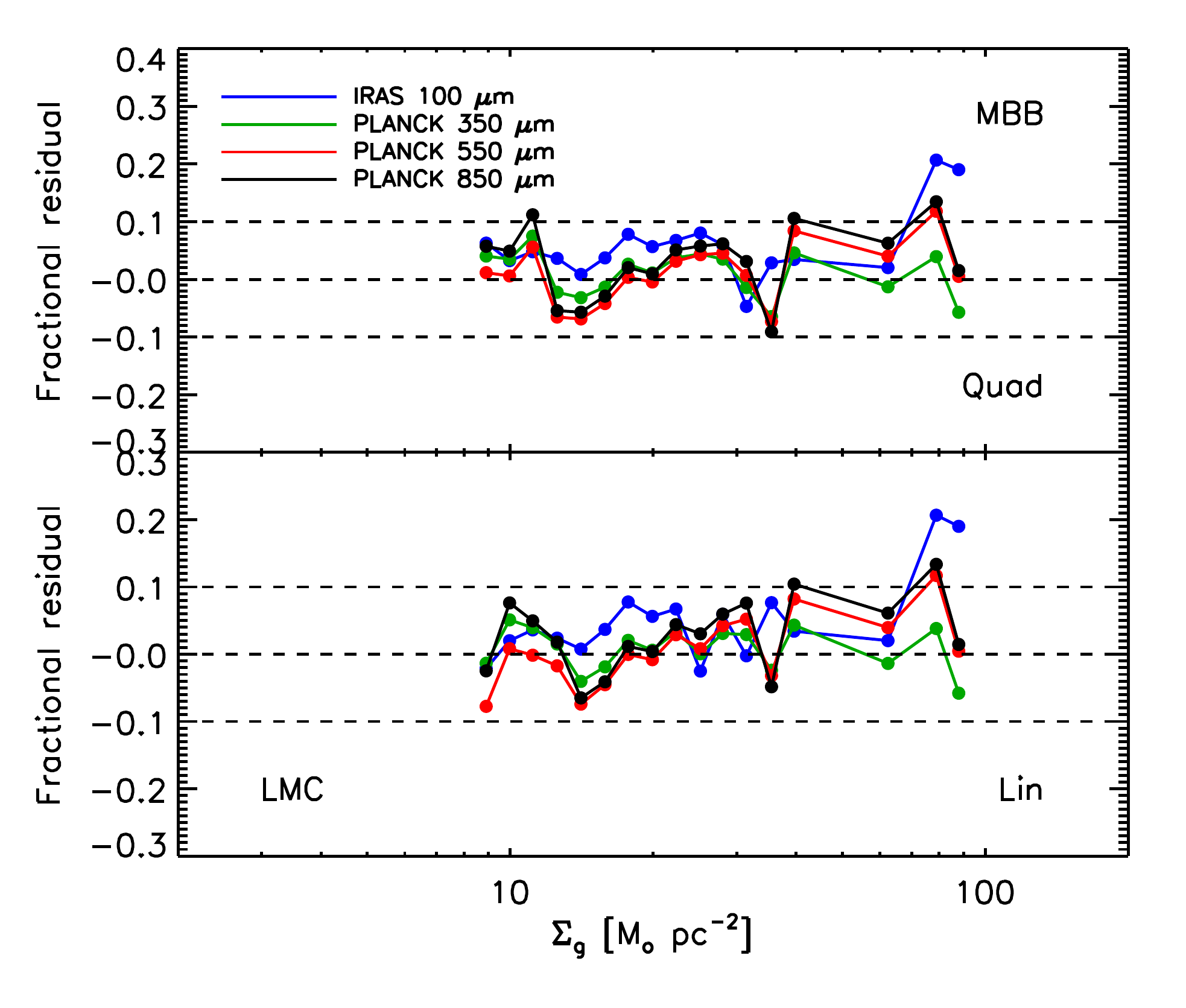}
\includegraphics[width=8cm]{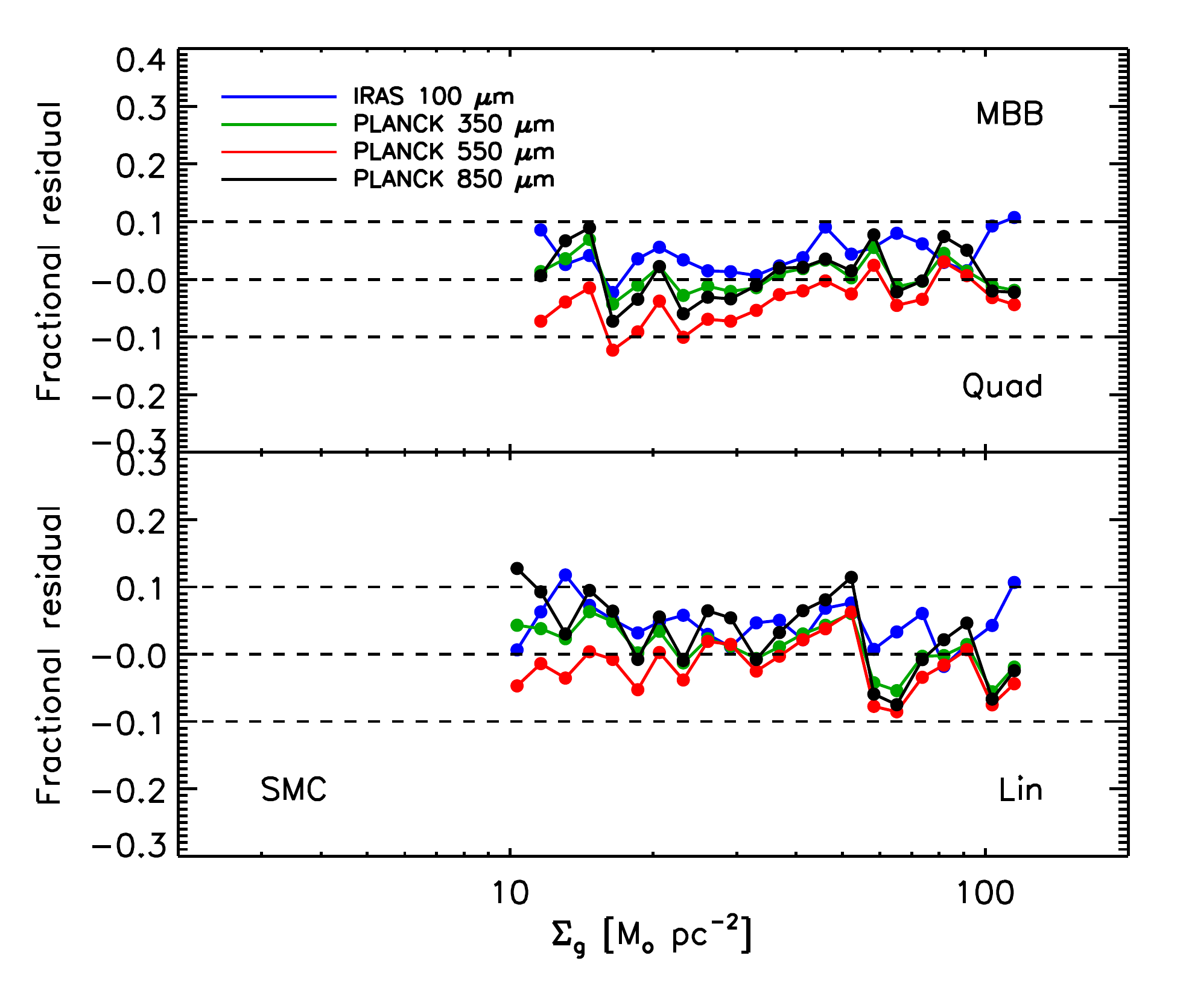}

\caption{Fractional residuals from the fits between the stacked dust SED at 100, 350, 550, arm 850 $\mu$m and the modified black body model described in Section \ref{modeling_section}, as a function of gas surface density, in the LMC (top) and SMC (bottom).}
\label{plot_residuals}
\end{figure}

\noindent Equations are similar for the other dust parameters. Random draws from the 1D PDFs, the best-fit model, and the data are shown in Figure \ref{plot_sed} for the $\Sigma_g$ $=$ 45.9 \Msu pc$^{-2}$) bin. We note that the parameter values that minimize the $\chi^2$ and the expectation values of each 1D PDF are consistent within errors. The SEDs and best-fit models are also shown in Figure \ref{plot_sed} for all the gas surface density bins in the LMC and SMC.\\
\indent The dust SED fitting is performed for gas surface density bins in which the mean dust emission is greater than the standard deviation in dust emission in this bin, or in other words, $S_{\mathrm{band}}^{\mathrm{mean}}$ $>$ $S_{\mathrm{band}}^{\mathrm{std}}$. The lowest gas surface density bin in which the dust can be modeled corresponds to $\Sigma_g$ $\sim$ 10 \Msu pc$^{-2}$ in both the LMC and SMC (see Figure \ref{plot_gas_dust_emission}). Additionally, we fit the dust SED for the dust emission obtained from a quadratic and linear estimation and subtraction of the Milky Way foreground cirrus emission.  \\
\indent The fractional residuals from the fits in the different bands are shown in Figure \ref{plot_residuals}, and lie within 10\%. This indicates that the dust SED in the different gas surface density bins is well modeled by a modified black body with a spatially varying (but spectrally constant) spectral emissivity index (the SMBB model). \\

\subsection{Self-calibration of the dust emissivity}\label{kappa_cal_section}

\indent The dust surface density and emissivity are degenerate, since the dust emission is proportional to their product. It is therefore important to calibrate the dust FIR emissivity using our fitting technique based on independent measurements of the dust surface density. Following \citet{gordon2014}, we self-calibrate $\kappa_0$ by applying our dust model to the Milky Way diffuse dust SED from \citet{compiegne2011}, for which the hydrogen column density is known and the dust-to-gas ratio can be estimated from depletions derived from UV absorption spectroscopy \citep{jenkins09}. Knowing the H/D (from depletions) and gas column density (from 21 cm emission) for this sight-line, we can derive the dust surface density independently from FIR emission. The dust emissivity is then estimated by dividing the FIR emission at 160 $\mu$m by the dust surface density. We choose to calibrate the emissivity at 160 $\mu$m 1) to minimize the potential temperature dependence sometimes seen in laboratory measurements of the opacity of dust analogs at longer wavelengths \citep{coupeaud2011}, and 2) to minimize the potential contamination by hot or stochastically heated dust. The resulting emissivity can finally be applied to the modeling of the LMC and SMC dust SED. \\
\indent The diffuse MW SED in \citet{compiegne2011}, which corresponds to an \his column density $N($\hi$)$ $=$ 10$^{20}$ cm$^{-2}$, has values 0.70 MJy sr$^{-1}$, 0.55 MJy sr$^{-1}$,  0.20 MJy sr$^{-1}$, and 0.052 MJy sr$^{-1}$ for the 100 $\mu$m DIRBE band and 350 $\mu$m, 550 $\mu$m, 850 $\mu$m Planck bands. We have converted these values by removing the 0.77 correction for ionized gas, since the spectroscopic measurement of depletions does not include the ionized gas contribution. The corresponding uncertainties are 0.082 MJy sr$^{-1}$, 0.048 MJy sr$^{-1}$, 0.033 MJy sr$^{-1}$, and  0.011 MJy sr$^{-1}$ for $N($\hi$)$ $=$ 10$^{20}$ cm$^{-2}$. The \his column density toward this Galactic sight-line is $<$ 5.5$\times10^{20}$ cm$^{-2}$, corresponding to a median depletion factor $F_{*}$ $=$ 0.4 \citep[see also][]{gordon2014}. In the Milky Way, a depletion factor $F_{*}$ $=$ 0.4 corresponds to a hydrogen-to-dust ratio of $(H/D)_{\mathrm{cal}}$ $=$ 150, as obtained from the depletion patterns in \citet{jenkins09}. \\
\indent  We run this SED normalized to N(H) $=$ 10$^{20}$ cm$^{-2}$ (corresponding to a hydrogen surface density $\Sigma_{\mathrm{cal}}$(\hi) $=$ 0.8 \Msu pc$^{-2}$) through our dust fitting code with a dummy dust emissivity of 1 cm$^{-2}$ g$^{-1}$ at 160 $\mu$m and derive a dummy dust surface density of 6.6$\times10^{-2}$ \Msu pc$^{-2}$. Since we know that the dust surface density for this SED must be $\Sigma_d^{\mathrm{cal}}$ $=$ $\Sigma_{\mathrm{cal}}$(\hi)/$(H/D)_{\mathrm{cal}}$ $=$ 0.8/150 $=$ 5.3$\times10^{-3}$ \Msu pc$^{-2}$, the dust emissivity must be $\kappa_0$ $=$ 12.4 cm$^{2}$ g$^{-1}$ at $\lambda_0$ $=$ 160 $\mu$m. The error on $\kappa_0$ is 50\% and is dominated by errors on the dust surface density from the SED fitting. This value is consistent with different physical dust models, such as \citet{weingartner2001} ($\kappa_{160}$ $=$ 9.97 cm$^2$ g$^{-1}$), \citet{zubko2004} ($\kappa_{160}$ $=$ 10.75---15 cm$^2$ g$^{-1}$ depending on the fraction of graphite and amorphous carbon), \citet{draine2007,draine2014} ($\kappa_{160}$ $=$ 12.5 cm$^2$ g$^{-1}$), but is lower than the value of 30.2 cm$^2$ g$^{-1}$ derived in \citet{gordon2014} and \citet{gordon2017}. The difference might be explained by the difference in the modeling approach and the bands used. For instance, we include measurement errors, while \citet{gordon2014} do not. We also are using four IRAS and PLANCK between 100 $\mu$m and 850 $\mu$m, while \citet{gordon2014} use five Herschel bands between 100 $\mu$m and 500 $\mu$m. Using our fitting code with the same inputs as \citet{gordon2014}, i.e, the same Herschel 5-band SED, the same calibration covariance matrix, excluding measurement errors, we find the exact same emissivity at 160 $\mu$m (30.2 cm$^2$ g$^{-1}$), which validates the robustness of our fitting algorithm. \\
\indent Applying the emissivity calibrated in the MW to the LMC and SMC observations assumes that the dust emissivity is similar between the MW, LMC, and SMC, which may not be correct. Indeed, \citet{meixner2010} and \citet{galliano2011} find evidence that the FIR emissivity of dust differs between these three galaxies. However, most of the difference in these studies lies in the spectral emissivity index, which we leave as a free parameter. Nevertheless, we also calibrate $\kappa_{160}$ based on the LMC and SMC FIR SEDs and depletion measurements corresponding to $\Sigma_g$ $=$ 22 \Msu pc$^{-2}$ in the LMC and  $\Sigma_g$ $=$ 65 \Msu pc$^{-2}$ in the SMC. These gas surface densities are chosen to be in the atomic regime, since the HI-H$_2$ transitions occur at $\Sigma_g$ $=$ 45 \Msu pc$^{-2}$ in the LMC and $\Sigma_g$ $=$ 110 \Msu pc$^{-2}$ in the SMC \citep{RD2014}, and to minimize the measurement errors. The SEDs corresponding to those gas surface densities, shown in Figure \ref{plot_gas_dust_emission}, are listed in Table 2 for the LMC and SMC. \citet{tchernyshyov2015} compiled depletion measurements in the LMC and SMC. The H/D estimates in \citet{tchernyshyov2015} (their Figure 10) are calculated as a function of $F_{*}$, a parameter that describes the collective behavior of depletions for all elements, and is roughly correlated with mean density. To translate the variations of H/D vs $F_{*}$ into variations of G/D vs gas surface density, we 1) multiplied the H/D in \citet{tchernyshyov2015} by 1.36 to account for Helium, and 2) examined the correlation between $F_{*}$ and $\Sigma_g$ and found that they are linearly correlated (with significant scatter). In the LMC F$_*$---$\Sigma_g$ relation, $F_{*}$ $=$ 0 roughly corresponds to $\Sigma_g$ $\sim$ 7 \Msu pc$^{-2}$, while $F_{*}$ $=$ 1 corresponds to $\Sigma_g$ $\sim$ 43 \Msu pc$^{-2}$. In the SMC,  $F_{*}$ $=$ 0 corresponds to $\Sigma_g$ $\sim$ 6 \Msu pc$^{-2}$, while $F_{*}$ $=$ 1 corresponds to $\Sigma_g$ $\sim$ 100 \Msu pc$^{-2}$. Hence, assuming a linear relation between $\Sigma_g$ and $F_*$, the surface density of $\Sigma_g$ $=$ 22 \Msu pc$^{-2}$ chosen to calibrate $\kappa_{160}$ in the LMC corresponds to $F_*$ $=$ 0.42 and $\Sigma_g$ $=$ 65 \Msu pc$^{-2}$ in the SMC corresponds to $F_*$ $=$ 0.80. The corresponding depletion-based G/D estimates are $(G/D)_{\mathrm{dep}}$ $=$ 328 in the LMC and $(G/D)_{\mathrm{dep}}$ $=$ 800 in the SMC. Combining the estimated G/D, $\Sigma_g$, and FIR SEDs, we derive $\kappa_{160}$  $=$ 6.1$\pm$0.5 cm$^2$ g$^{-1}$ in the LMC and $\kappa_{160}$ $=$ 1.65$\pm$0.2 cm$^2$ g$^{-1}$ in the SMC. These values are lower than any estimates from dust models based on MW measurements and are consistent with the hypothesis that dust properties in the LMC and SMC differ substantially from the MW. Due to the discrepancy with physical dust models, we assume $\kappa_0$ $=$ 12.4 cm$^{2}$ g$^{-1}$ at $\lambda_0$ $=$ 160 $\mu$m (based on the calibration with the MW SED) in the following, but we explore the results based on the LMC/SMC calibration in Section \ref{comparison_depletions}. 

\begin{deluxetable*}{ccccccccc}
\tabletypesize{\scriptsize}
\tablecolumns{9}
\tablewidth{\textwidth}
\tablecaption{SEDs, gas and depletion measurements used to calibrate $\kappa_{160}$ in the LMC and SMC}
\tablenum{2}
 
 \tablehead{ & $S_{100}$ & $S_{350}$ & $S_{550}$ & $S_{850}$ & $\Sigma_g$ & $F_*$ & $(G/D)_{\mathrm{dep}}$ & $\kappa_{160}$ \\
 & MJy/sr & MJy/sr & MJy/sr & MJy/sr & \Msu pc$^{-2}$ & & & cm$^2$ g$^{-1}$ \\
  } \\
  
 \startdata
 &&&&&&&&\\
LMC & 22.4$\pm$16.5 & 8.2$\pm$4.1 & 2.9$\pm$1.4 &  0.87$\pm$0.42 & 22 & 0.42 & 328 & 6.1 \\
SMC & 13.6$\pm$1.05 & 5.95$\pm$0.43 & 2.37$\pm$0.16 & 0.80$\pm$0.051 & 65 & 0.80 & 800 & 1.65 \\
   \enddata
  \label{kappa_lmc_smc_table}

\end{deluxetable*}

\section{Results}\label{results_section}

\indent Before getting into a quantitative analysis of the variations of the dust abundance and properties, it is worth pointing out that Figure \ref{figs_3col} provides a clear visualization of the spatial variations of the dust abundance and properties. Bright star-forming regions appear dominantly blue/green, indicating a high ratio of bright FIR emission to \his surface density. The outskirts of the LMC and SMC and regions of low surface density  (i.e., holes in the galaxies) appear to be predominantly red, indicating that the ratio of \his surface density to FIR emission is high. The color variations in Figure \ref{figs_3col} can be explained by variations in the dust temperature (i.e., heating rate) and dust abundance (G/D). In this section, we quantify those variations using the stacking analysis and dust modeling described in Section \ref{method_section}. \\
\indent We derive the dust surface density, dust temperature, and spectral emissivity index from the dust modeling for each interval of gas surface density following the method outlined in Section \ref{method_section}. The variations of these parameters are described in the following sections. Integrating the dust and gas surface densities yield total dust and gas masses summarizes in Table 3 for both MW cirrus estimation and subtraction.  For comparison, \citet{gordon2014} find total dust masses of (7.3$\pm$1.7)$\times 10^5$ in the LMC and (8.3$\pm$2.1)$\times 10^4$ in the SMC for gas masses of 4$\times 10^8$ for both galaxies (adjusted from the 3$\times 10^8$ \Msu \his mass to account for Helium), corresponding to G/D values of 560 for the LMC and 4900 for the SMC. Hence, we find slightly lower dust masses and higher G/D, but consistent within the errors. 

\begin{deluxetable*}{ccccccc}
\tabletypesize{\scriptsize}
\tablecolumns{7}
\tablewidth{\textwidth}
\tablecaption{Total dust and gas masses in the LMC and SMC}
\tablenum{3}
 
 \tablehead{
  & \multicolumn{3}{c}{LMC} & \multicolumn{3}{c}{SMC}} \\
  
 \startdata

 & Gas & Dust & G/D & Gas & Dust & G/D \\
 & \Msun & \Msun &  & \Msun & \Msun & \\
 &&&&&& \\
 \hline
 &&&&&&\\
Quadratic fit to MW cirrus emission & (4.0$\pm$0.04)$\times 10^8$  & (5.1$\pm$0.47)$\times 10^5$  & 785  $\pm$ 72 & (4.7$\pm$0.04)$\times 10^8$ & (7.3$\pm$0.77)$\times 10^4$  & 6380 $\pm$ 680\\
Linear  fit to MW cirrus emission  & (4.0$\pm$0.04)$\times 10^8$  & (5.3$\pm$0.5)$\times 10^5$ & 758 $\pm$ 68 & (4.8$\pm$0.04)$\times 10^8$  & (6.7$\pm$0.7)$\times 10^4$ & 7070 $\pm$ 770\\

   \enddata
  \label{total_masses_table}
  \tablecomments{The total masses only include stacked pixels where a dust model can be fit, i.e., the standard deviation in the dust emission is less than the mean emission (see Section \ref{modeling_section}). Hence, the total gas masses quoted in this table depend slightly on the MW cirrus subtraction. We assume $\kappa_{160}$ $=$ 12.4 cm$^2$ g$^{-1}$ and $\alpha_{\mathrm{CO}}$ $=$ 6.4 \Msu pc$^{-2}$ (K km s$^{-1})^{-1}$ (LMC) and 21 \Msu pc$^{-2}$ (K km s$^{-1})^{-1}$ (SMC) to derive dust and gas masses, respectively.} 

\end{deluxetable*}

\subsection{Dust parameter variations with gas surface density}\label{dust_params_section}

\begin{figure*}
\includegraphics[width=8cm]{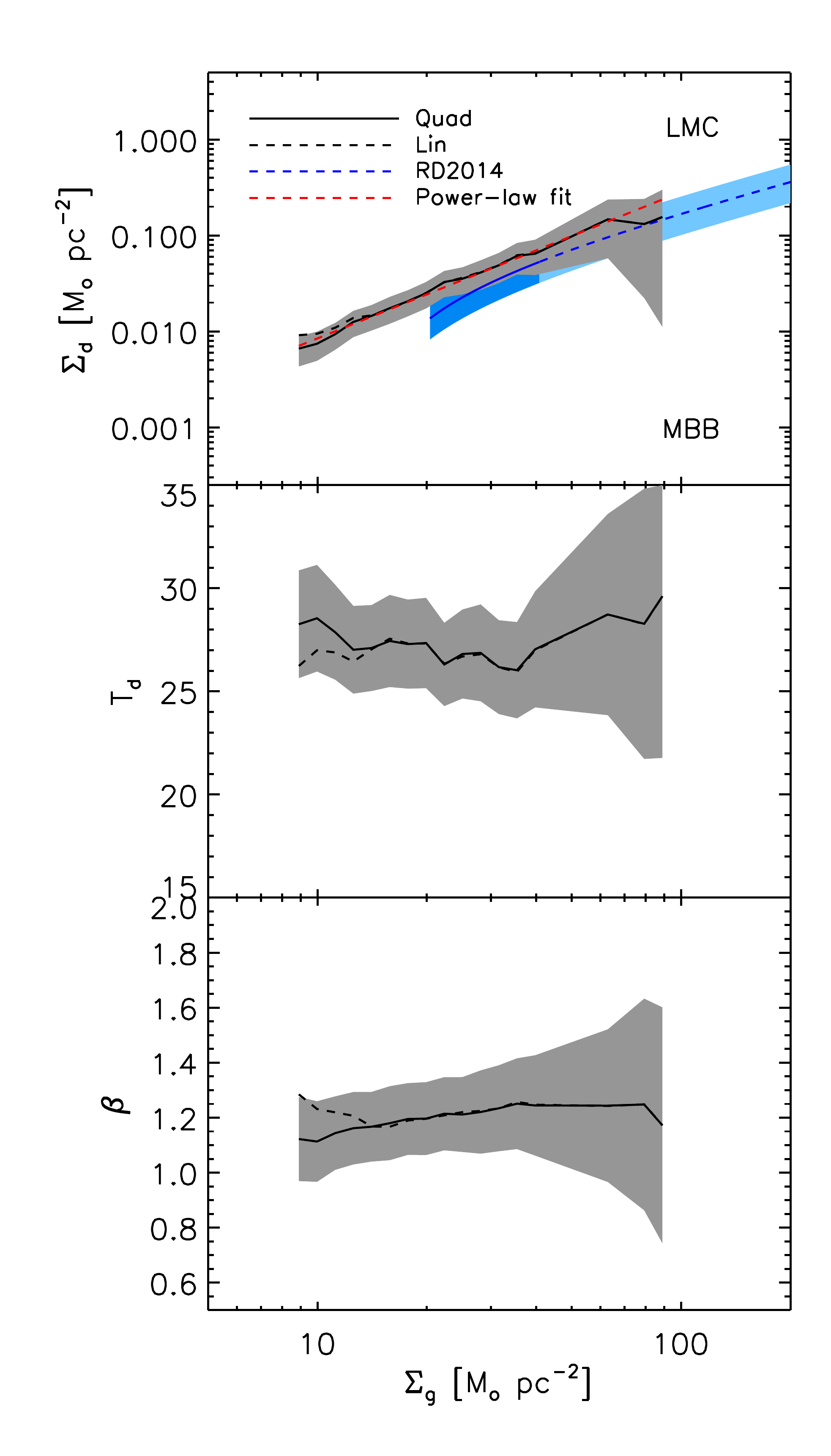}
\includegraphics[width=8cm]{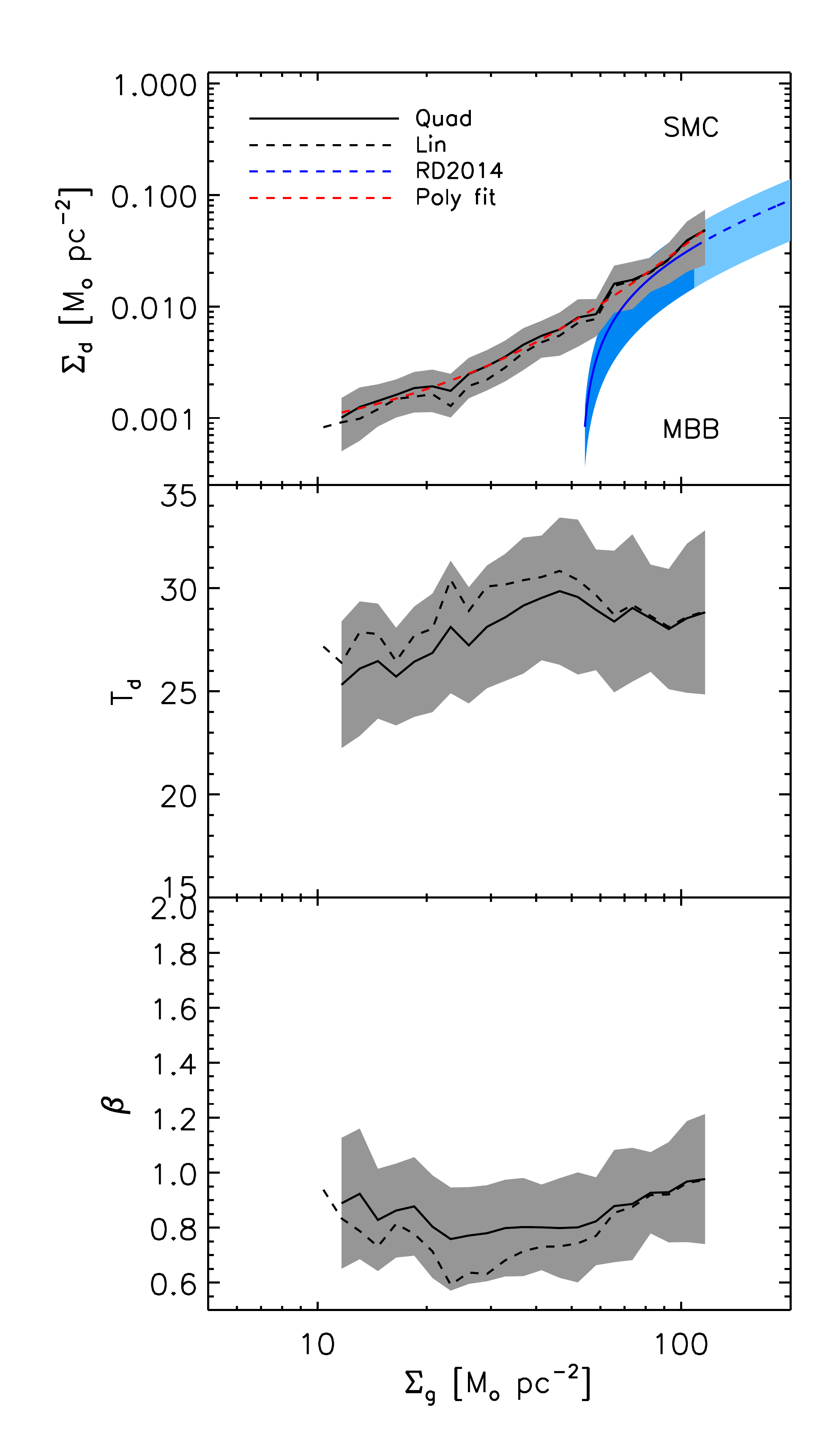}
\caption{Dust parameters ($\Sigma_d$, $T_d$, $\beta$) as a function of gas surface density, $\Sigma_g$ in the LMC (left) and SMC (right). The solid and dashed lines correspond to the dust emission after a quadratic and linear fit and subtraction of the MW cirrus, respectively. The shaded area represents the error, estimated at the 1$\sigma$ level. In the top panels, the blue solid line shows the linear relation between dust and gas derived in \citet{RD2014}, using \his surface densities $\Sigma$(\hi) $=$ 15---25 \Msu pc$^{-2}$ in the LMC and $\Sigma$(\hi) $=$ 40---80 \Msu pc$^{-2}$ in the SMC. The dark shaded area corresponds to the uncertainties on the \citet{RD2014} measurement (as linear fit with non-zero intercept to the dust-gas relation). The blue dashed line and light shaded area show the extrapolation of this relation. The red dashed line corresponds to a power-law fit (LMC) and polynomial fit (SMC) to the gas-dust relation.}
\label{plot_dust_params}
\end{figure*}

\indent  The resulting dust parameters are shown as a function of gas surface density in Figure \ref{plot_dust_params}. In both the LMC and SMC, the dust surface density increases non-linearly with gas surface density, indicating that the G/D is not constant with surface density. For comparison, a linear relation given by the slope (H/D) and intercept ($\Sigma_I$) quoted in \citet{RD2014} (H/D $=$ 380$_{-130}^{+250}$, $\Sigma_I$ $=$ 9.8 \Msu pc$^{-2}$ in the LMC and H/D $=$ 1200$_{-420}^{+1600}$, $\Sigma_I$ $=$ 39 \Msu pc$^{-2}$ in the SMC) is shown. The H/D in \citet{RD2014} is derived in the atomic ISM by fitting a linear function to the relation between gas and dust for $\Sigma_H$ $=$ 15---25 \Msu pc$^{-2}$ in the LMC and $\Sigma_H$ $=$ 40---80 \Msu pc$^{-2}$ in the SMC, and does not include the contribution of Helium. Thus, we multiply the H/D in \citet{RD2014} by 1.36 to compare to the trends of G/D derived here, which include Helium in the gas surface density. Overall, the relation between dust and gas surface densities derived here is in good agreement with the linear fit derived in \citet{RD2014} over the range where the analysis overlap. As noted before, this study allows us to derive the dust content of the LMC and SMC down to substantially lower surface densities, as shown in the top panels of Figure \ref{plot_dust_params}. Given that we perform a different cirrus estimation and subtraction and utilize different bands and a different fitting code, this agreement is a good test of the robustness of the results. \\
\indent The SMC contains $\sim$10 times less dust than the LMC for a given gas surface density. A linear scaling with metallicity between the LMC \citep[$Z = 0.5 Z_{\odot}$][]{russell92} and SMC \citep[$Z = 0.2 Z_{\odot}$][]{russell92} would predict a factor of 2.5 difference in G/D between the LMC and SMC. This suggests that the G/D varies non-linearly with metallicity, and is consistent with the metallicity-dependence of a large sample of dwarf galaxies presented in \citet{remyruyer2014}. Chemical evolution models such as \citet{asano2013, feldmann2015} have proposed explanations for this evolution: above a certain critical metallicity ($Z_{\mathrm{crit}}$ $=$ 0.1---0.2 $Z_{\odot}$ for typical star formation timescales of 0.5-5 Gyr), dust formation in the ISM is efficient enough to balance the effects of dilution by galactic inflows and outflows. Below this critical metallicity, the dust input is dominated by the non-efficient formation process in evolved stars (AGB, SN), resulting in a low dust-to-metal ratio and subsequently a low dust abundance. These models predict that, for a given star-formation timescale in the range of 0.5---5 Gyr consistent with literature \citep{bigiel2008, bolatto2011, jameson2016}, the G/D in the LMC should be a factor 10 lower than in the SMC, consistent with our results. \\
\indent The dust temperature occupies a small range in the LMC and SMC, with values between 25 and 30 K. There is no apparent correlation between dust temperature and gas surface density on 75 pc scales (corresponding to 5' resolution for our dataset). Detailed 3D radiative transfer modeling would be necessary to interpret this result. \\
\indent The spectral emissivity index $\beta$ appears constant within the errors across the gas surface density range, with values within 0.8---1 in the SMC and 1.2---1.4 in the LMC. Such low values of $\beta$ indicate an amorphous carbon dust grain composition \citep{mennella1998, boudet2005}, a conclusion also reached by \citet{galliano2011} and  \citet{chastenet2017}. In both galaxies, there is no significant correlation between $\beta$ and surface density, indicating a relatively constant ratio of amorphous carbons to silicates.

\subsection{Dust abundance vs gas surface density}

\begin{figure*}
\centering
\includegraphics[width=8cm]{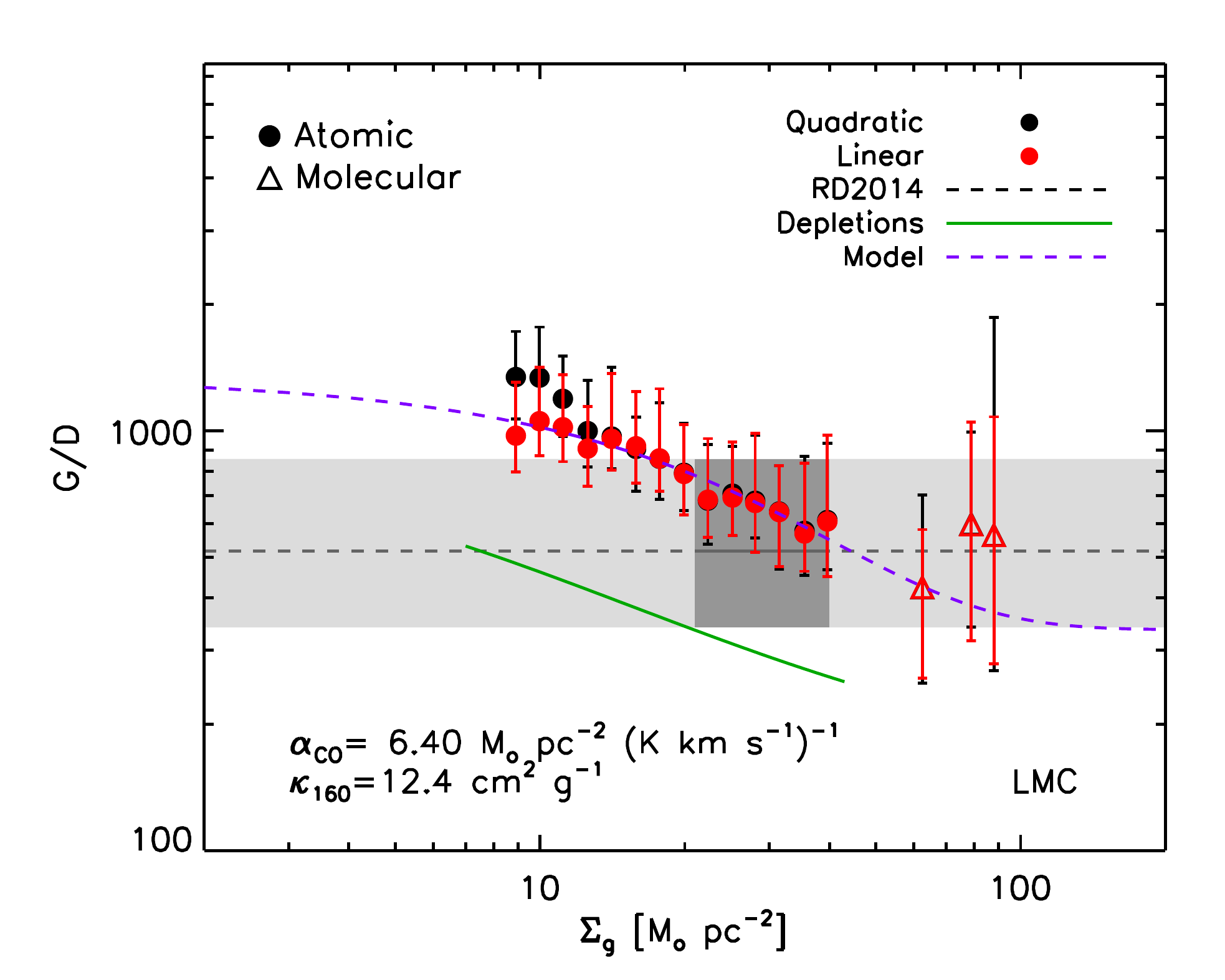}
\includegraphics[width=8cm]{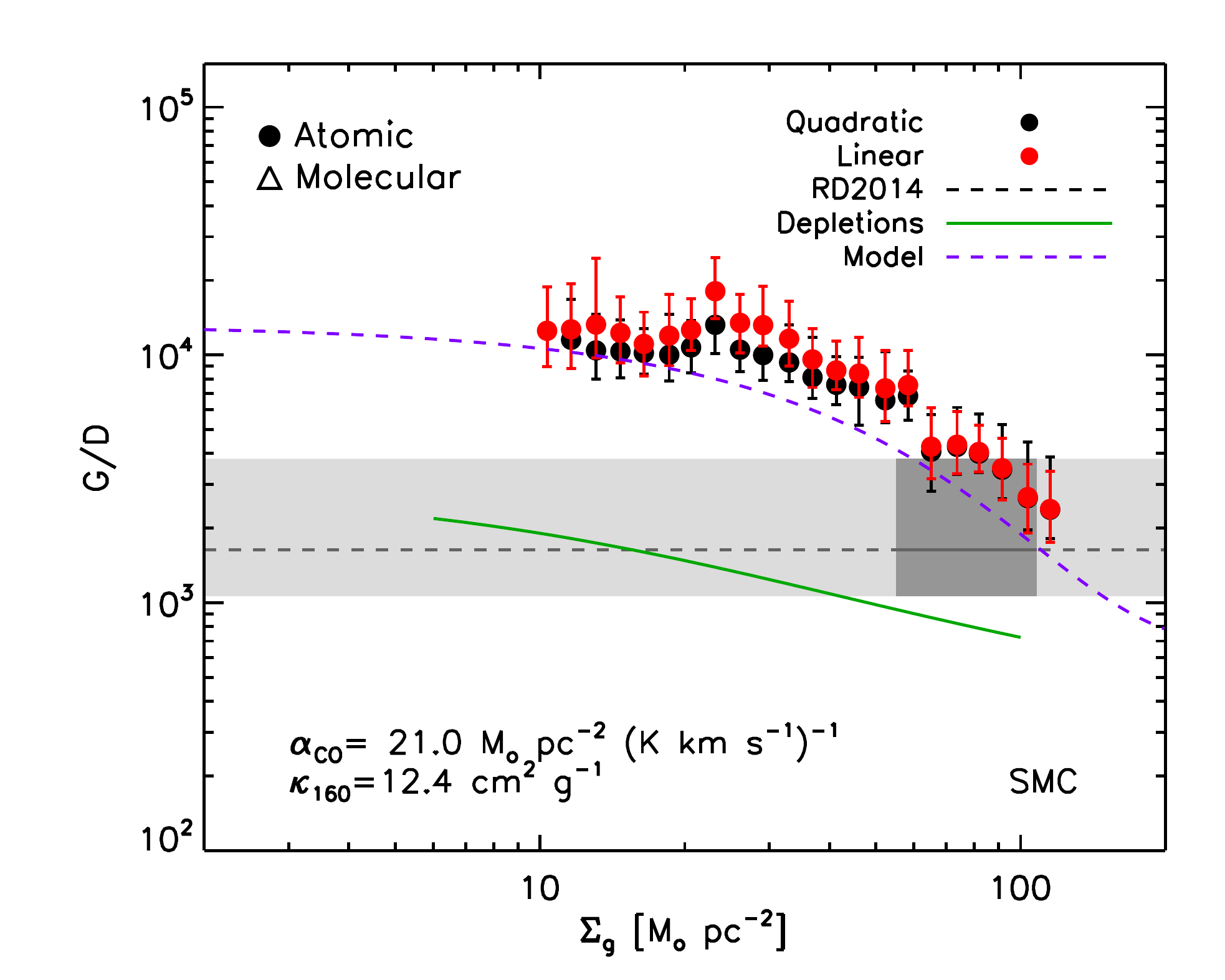}
\includegraphics[width=8cm]{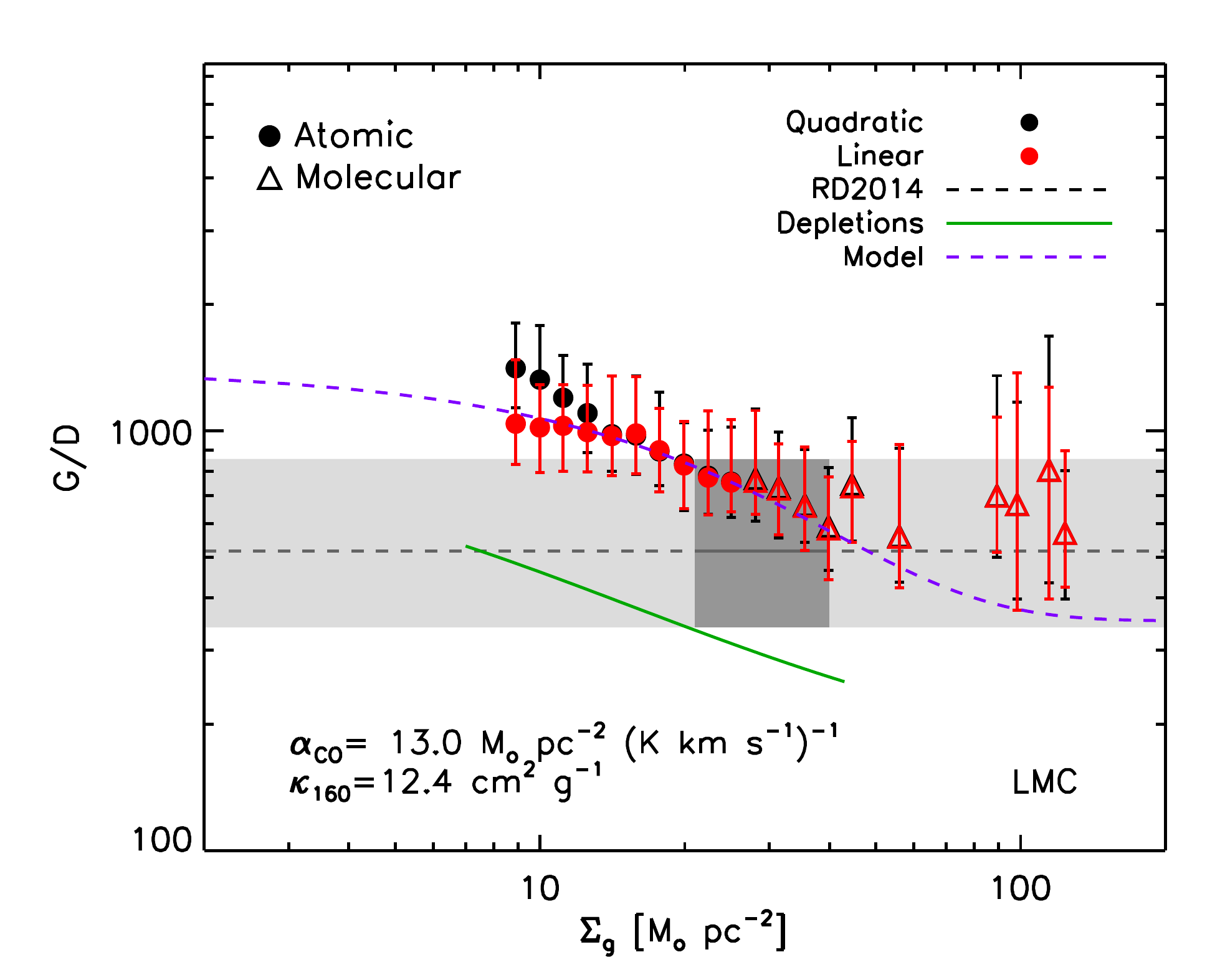}
\includegraphics[width=8cm]{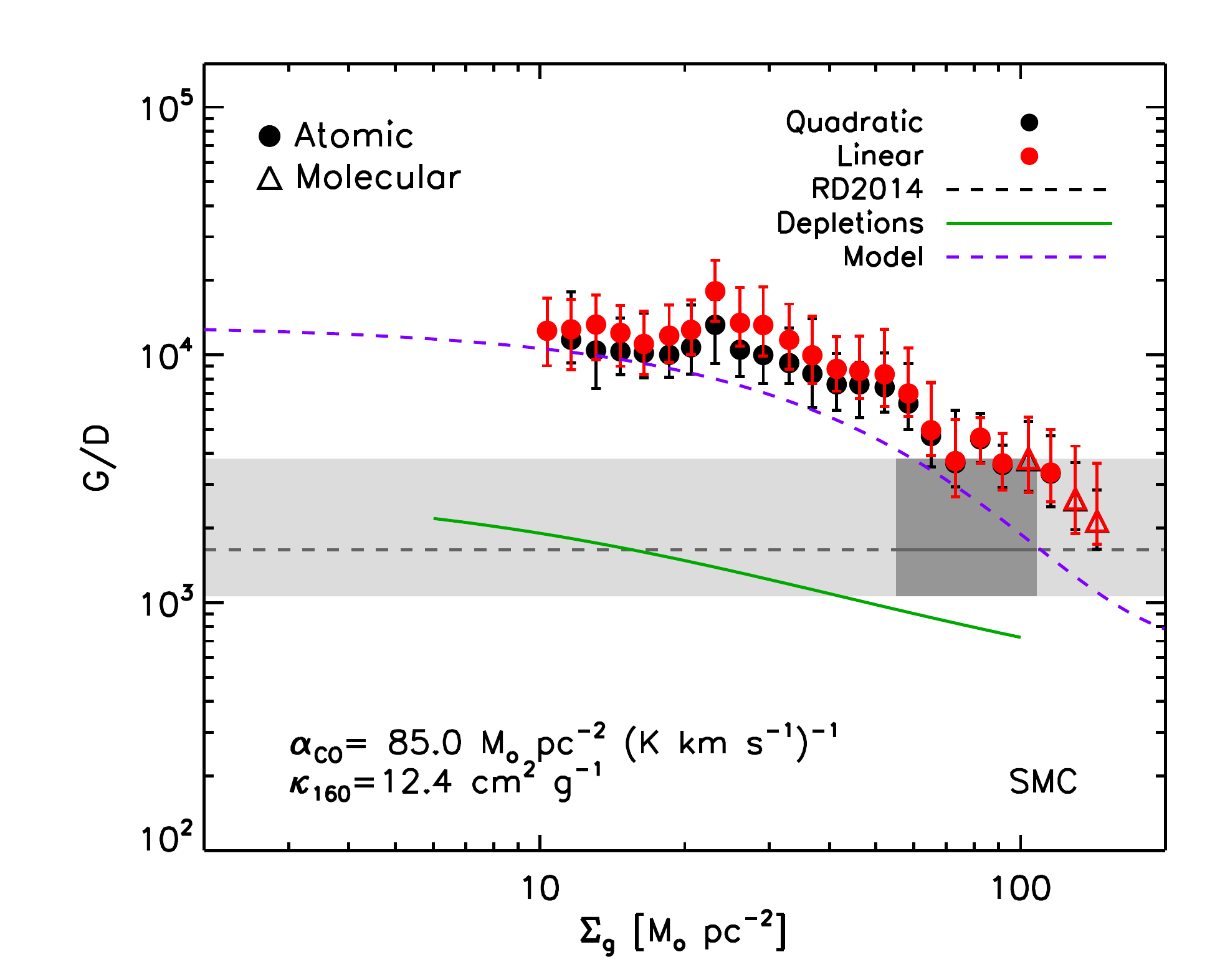}

\caption{Gas-to-dust ratio G/D stacked by 0.05 dex intervals in gas surface density in the LMC (left) and SMC (right). Error-bars correspond to the 50\% confidence interval. The top and middle panels correspond to different values of the CO-to-H$_2$ conversion factor, $\alpha_{\mathrm{CO}}$. In the top panels, $\alpha_{\mathrm{CO}}$ $=$ 6.4 \Msu pc$^{-2}$ (K km s$^{-1})^{-1}$ (1.5$\times$ Galactic, LMC) and 21 \Msu pc$^{-2}$ (K km s$^{-1})^{-1}$ (5$\times$ Galactic, SMC), and $\kappa_{160}$ $=$ 12.4 cm$^{2}$ g$^{-1}$, which are our fiducial values. In the bottom panels, our fiducial $\kappa_{160}$ $=$ 12.4 cm$^2$ g$^{-1}$ is assumed, but we assume the maximum possible value of $\alpha_{\mathrm{CO}}$ from \citet{RD2014}, or $\alpha_{\mathrm{CO}}$ $=$ 13 \Msu pc$^{-2}$ (K km s$^{-1})^{-1}$ (3$\times$ Galactic, LMC) and 85 \Msu pc$^{-2}$ (K km s$^{-1})^{-1}$ (20$\times$ Galactic, SMC). In all panels, circles and triangles correspond to the  atomic gas and molecular gas ($>$10\% molecular fraction). The black and red points correspond to the LMC/SMC dust emission after quadratic and linear fits and removal of the cirrus emission, respectively. The black solid line and dark shaded area show the constant G/D derived in \citet{RD2014} and its uncertainties, using \his surface densities $\Sigma$(\hi) $=$ 15---25 \Msu pc$^{-2}$ in the LMC and $\Sigma$(\hi) $=$ 40---80 \Msu pc$^{-2}$ in the SMC. The black dashed line and lightly shaded area show the extrapolation of this relation. The green line shows the G/D derived from depletions from UV absorption spectroscopy. The purple line corresponds to the analytical model developed in Section \ref{model_section} }
\label{gd_trends}
\end{figure*}

\indent The gas-to-dust ratio (G/D) is computed as the ratio of the gas surface density to the dust surface density $\Sigma_g$/$\Sigma_d$. The resulting G/D is shown as a function of gas surface density in Figure \ref{gd_trends}, for the case of a linear and quadratic Milky Way cirrus emission estimation and subtraction. Errors are computed by propagating the error on $\Sigma_d$ described in Section \ref{modeling_section} to the G/D measurement via a Monte Carlo simulation. The errors shown in Figure \ref{gd_trends} correspond to the 50\% confidence interval. We note that errors on the SED fits are dominated by calibration errors quoted in \citet{r3planck2015}, which, although not very large (10\% for IRAS, 6\% for Planck), can lead to substantial uncertainties on the dust temperature and therefore surface density due to the non-linear nature of the Planck function. Given the smoothness of the trends compared to the size of the error bars, it is likely that the errors are over-estimated.  \\
\indent We separate the gas surface density bins in two categories, ''atomic'' and ''molecular'', which corresponds to regions where molecular gas fraction as traced by CO emission is lower and greater than 10\%, respectively. In the SMC at 5' resolution, molecular gas clumps are completely diluted and do not appear in the figure (i.e., over 90 pc scales, the molecular gas fraction is always $<$ 10\%). This is expected due to the low metallicity of this galaxy and subsequent lack of dust shielding, leading to molecular gas filling factor that is substantially smaller than in the MW and LMC. In Figure \ref{gd_trends}, we also show the G/D vs gas surface density for the maximum possible value of $\alpha_{\mathrm{CO}}$ from \citet{RD2014}, or $\alpha_{\mathrm{CO}}$ $=$ 13 \Msu pc$^{-2}$ (K km s$^{-1})^{-1}$ (3$\times$ Galactic, LMC) and 85 \Msu pc$^{-2}$ (K km s$^{-1})^{-1}$ (20$\times$ Galactic, SMC). The choice of CO-to-H$_2$ conversion factor does not affect the trends observed here, understandably because the molecular gas represents such a small fraction of the gas mass.  \\
\indent For comparison, the G/D derived in \citet{RD2014} is shown in Figure \ref{gd_trends}. The G/D values derived here (black and red points) and in \citet{RD2014} (gray band) are consistent within errors in the limited gas surface density range probed by the measurements in \citet{RD2014}. The G/D presented here is slightly higher than the value in \citet{RD2014} for both galaxies. However, we will see in Section \ref{gd_slope_ratio} that this discrepancy is due to the difference in measurement technique.\citet{RD2014} derived the G/D as the slope (derivative) of the dust-gas relation, assumed to be linear, while we simply compute the ratio of the gas to dust surface densities in Figure \ref{gd_trends}. \\
\indent Over the dynamic range observed in the LMC and SMC ($\Sigma_g$ $\sim$ 10---100 \Msu pc$^{-2}$, shown by pink contours in Figure \ref{hi_maps}), the dust abundance (G/D) varies by a factor $\sim$ 3, from 1500 to 500 in the LMC, and by a factor $\sim$7 in the SMC from 1.5$\times$10$^4$ down to 2000. The trends are robust against the method implemented to estimate and subtract the foreground cirrus MW dust emission. Moreover, the atomic-to-molecular gas transition occurs at $\Sigma_g$ $=$ 45 \Msu pc$^{-2}$ in the LMC and $\Sigma_g$ $=$ 110 \Msu pc$^{-2}$ in the SMC \citep{RD2014}. Therefore, most of the G/D variations observed here occur in the atomic ISM. The origins and implications of these trends are discussed in Section \ref{discussion_section}. \\
\indent We note that the gas surface density, $\Sigma_g$, is present in the numerator of both the abscissa and ordinate in Figure \ref{gd_trends}. However, Figure \ref{gd_trends} shows an anti-correlation between G/D and gas surface density. If the trend were due to a degeneracy between these two dependent variables, we would expect a correlation. This implies that the observed decrease of G/D with surface density is not a result of noise.

\subsection{Comparison to elemental depletions}\label{comparison_depletions}

\begin{figure*}
\centering

\includegraphics[width=8cm]{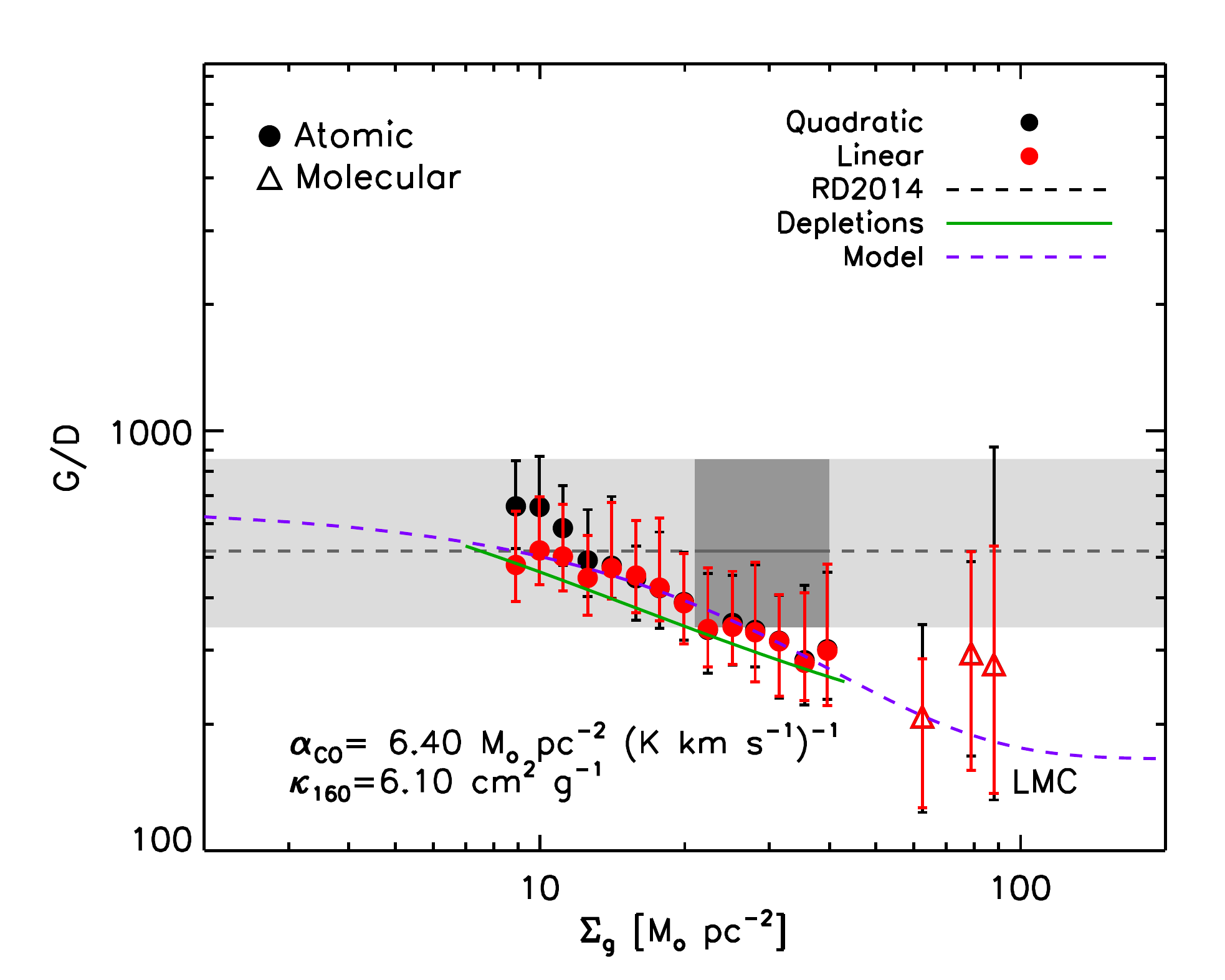}
\includegraphics[width=8cm]{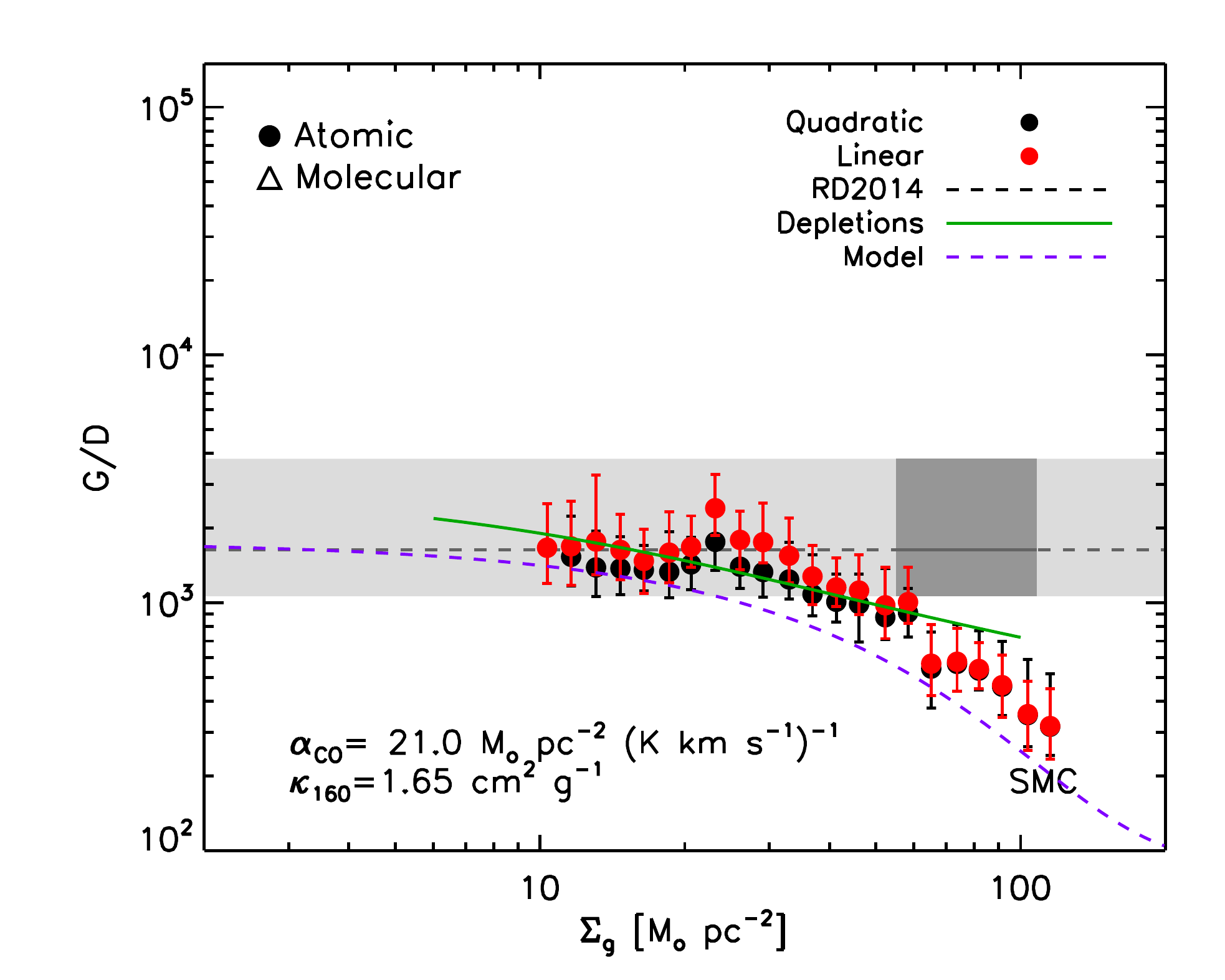}
\caption{Gas-to-dust ratio G/D stacked by 0.05 dex intervals in gas surface density in the LMC (left) and SMC (right). Error-bars correspond to the 50\% confidence interval. We assume $\alpha_{\mathrm{CO}}$ $=$ 6.4 \Msu pc$^{-2}$ (K km s$^{-1})^{-1}$ (1.5$\times$ Galactic, LMC) and 21 \Msu pc$^{-2}$ (K km s$^{-1})^{-1}$ (5$\times$ Galactic, SMC) and a $\kappa_{160}$ opacity value derived from calibrating the G/D toward an LMC/SMC sight-line with known SED and depletions. Circles and triangles correspond to the  atomic gas and molecular gas ($>$10\% molecular fraction). The black and red points correspond to the LMC/SMC dust emission after quadratic and linear fits and removal of the cirrus emission, respectively. The black solid line and dark shaded area show the constant G/D derived in \citet{RD2014} and its uncertainties, using \his surface densities $\Sigma$(\hi) $=$ 15---25 \Msu pc$^{-2}$ in the LMC and $\Sigma$(\hi) $=$ 40---80 \Msu pc$^{-2}$ in the SMC. The black dashed line and lightly shaded area show the extrapolation of this relation. The green line shows the G/D derived from depletions from UV absorption spectroscopy. The purple line corresponds to the analytical model developed in Section \ref{model_section} }
\label{gd_trends_other_kappa}
\end{figure*}

\indent The G/D calculated from depletions \citep{tchernyshyov2015} is shown in Figure \ref{gd_trends} (dashed green line). The depletion measurements are converted from H/D to G/D as explained in Section \ref{kappa_cal_section}.  \\
\indent The slope of the G/D dependence on surface density in the depletion measurements is consistent with the slope derived here. However, the G/D derived from FIR emission and from depletions are offset by a factor of $\sim$2 in the LMC, and $\sim$8 in the SMC. A factor of 2 could easily be explained by differences in the dust properties, particularly in the FIR dust emissivity, between the LMC and the MW, where $\kappa_{160}$ is calibrated. However, a factor of 8 in the SMC is more difficult to explain with only emissivity variations. As shown in Section \ref{kappa_cal_section}, reconciling the G/D measured from FIR emission with the G/D estimated from depletions would require $\kappa_{160}$ $=$ 6.1 cm$^2$ g$^{-1}$ in the LMC and $\kappa_{160}$ $=$ 1.65 cm$^2$ g$^{-1}$ in the SMC. This is shown in Figure \ref{gd_trends_other_kappa}, which shows the G/D variations in the LMC and SMC assuming those values of $\kappa_{160}$. In this case, the G/D variations with gas surface density from FIR emission and depletions agree fairly well. \\
\indent While the LMC opacity value is in reasonable agreement with dust models \citep[e.g., ][]{draine2007, draine2014, zubko2004}, the opacity calibrated on the SMC SED and depletion measurement seems unreasonably low compared to models in the literature. However, a plausible explanation comes from the comparison of the \his column density derived from UV absorption spectroscopy and from 21 cm emission in the LMC and SMC. \citet{welty2012} find that \his column densities in the SMC derived from 21 cm emission are a factor 2---10 higher than column densities derived from Lyman $\alpha$ damping wing fitting in the surface density range probed by depletions ($\Sigma$(\hi) $<$ 40 \Msu pc$^{-2}$). They explain this discrepancy by the complex spatial and kinematic structure of the SMC, leading to a substantial amount of \his being located behind the stars inducing the absorption. In the LMC, the discrepancy is lower than a factor 2 owing to the much thinner structure of the disk. Thus, the differences in \his surface density between UV absorption and 21 cm measurements are consistent with the discrepancy in G/D between depletion and FIR/21 cm emission measurements observed here.

\section{Discussion}\label{discussion_section}

\subsection{G/D as a ratio or a slope}\label{gd_slope_ratio}

\indent In \citet{RD2014}, we emphasized the difference between computing G/D as a ratio ({\it integrated measurement}) and as a the slope of the relation between dust and gas surface densities ({\it local} measurement). In Figure \ref{gd_trends}, there is an apparent disagreement between the G/D measurement in \citet{RD2014} (gray band) and the study presented here (black and red points), with the G/D derived here being a factor of a few higher than in \citet{RD2014}. However, the G/D in Figure \ref{gd_trends} is computed as a direct ratio $\Sigma_g$/$\Sigma_d$, while the \citet{RD2014} G/D measurement shown in gray is based on a linear fit (with a non-zero intercept) to the relation between dust and gas surface densities. As suggested by the top panels of Figure \ref{plot_dust_params}, showing the relation between dust and gas surface densities from this analysis, the dust content of the LMC and SMC derived here is in very good agreement with the dust content observed in \citet{RD2014}. However, since the gas and dust surface densities have a non-linear (in fact super-linear) relation, their ratio is significantly different from their derivative. \\
\indent In order to derive the G/D as the derivative of the dust-gas relation as in \citet{RD2014}, we fit the relation between dust and gas in the top panels of Figure \ref{plot_dust_params} with different functional forms. For the LMC, the log gas-log dust relation can be approximated by a linear function (Figure \ref{plot_dust_params}). Therefore, a power-law (a linear fit in log-space) provides an excellent fit, which is shown as a red dashed line in Figure \ref{plot_dust_params}. The resulting fit in the LMC is:

\begin{equation}
\Sigma_d^{\mathrm{fit}} = 2.5 \times 10^{-4} \Sigma_g^{1.5}
\label{lmc_fit}
\end{equation}

\noindent In the SMC, the log gas-log dust relation exhibit some curvature, which is not captured by a power-law fit. In this case, we find that a polynomial of order 3 provides the best fit to the relation between dust and gas (also shown as a red dashed line in Figure \ref{plot_dust_params}). The resulting relation is:

\begin{multline}
 \Sigma_d^{\mathrm{fit}} = 4.87  \times 10^{-4}  + 4.19  \times 10^{-5} \Sigma_g  \\
 + 8.18 \times 10^{-7} \Sigma_g^2 + 2.06  \times 10^{-8} \Sigma_g^3
\label{smc_fit}
\end{multline}

\begin{figure*}
\centering
\includegraphics[width=8cm]{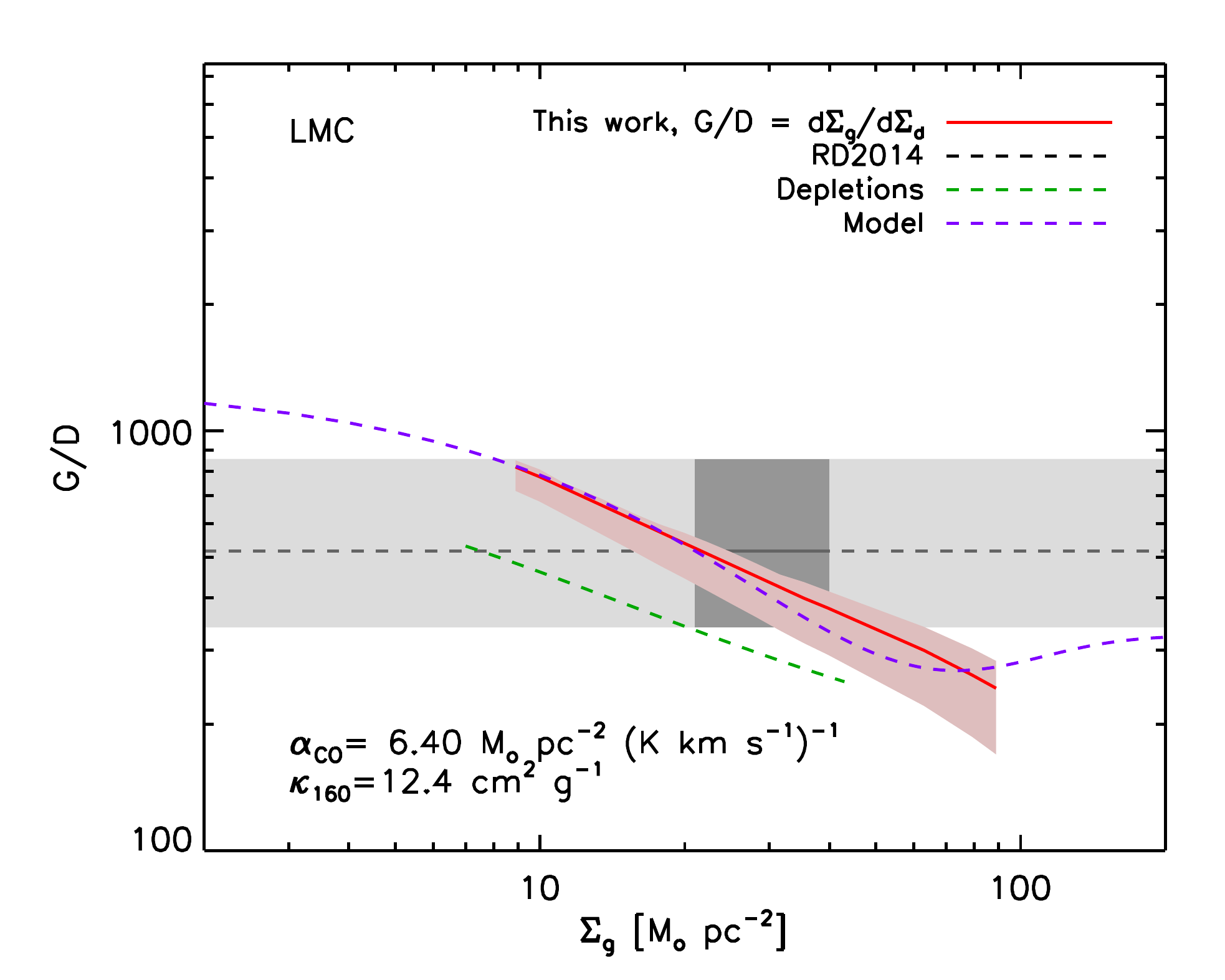}
\includegraphics[width=8cm]{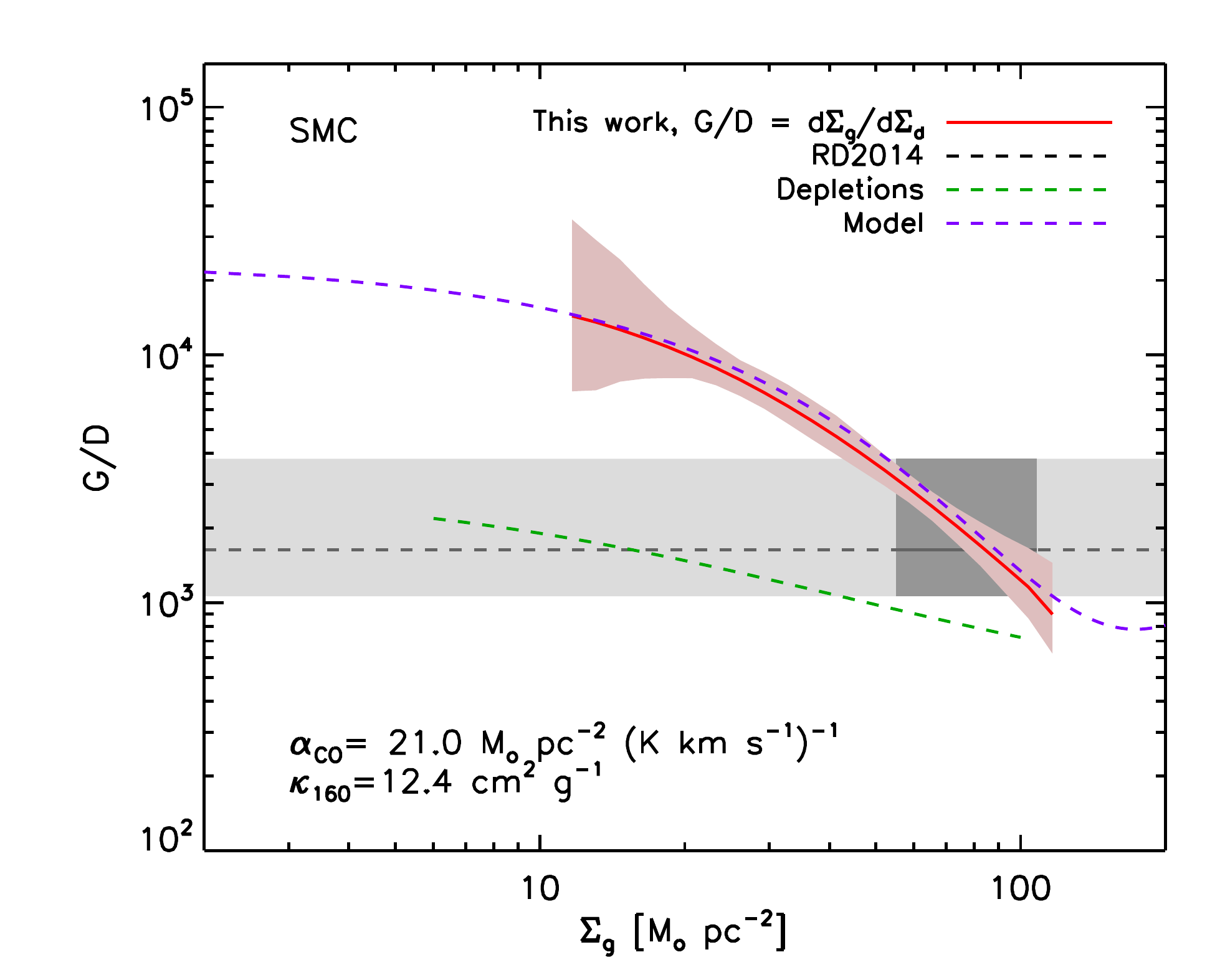}
\caption{Gas-to-dust ratio G/D derived as the derivative of the best-fit relation between dust and gas (see Figure \ref{plot_dust_params}. Error-bars correspond to the 50\% confidence interval. For the LMC, we fit a power-law to the dust-gas relation (red line in the top panel of Figure \ref{plot_dust_params} and Equation \label{lmc_fit}. In the SMC, we fit a polynomial to the dust gas relation, given by Equation \ref{smc_fit}, shown by the red line in Figure \ref{plot_dust_params}. The band corresponds to the 1$\sigma$ uncertainties. The dark gray band shows the G/D slope derived in \citet{RD2014}. The light gray band shows the extrapolation of this constant G/D to all surface densities. The purple dashed line shows the simple analytical model for accretion of gas-phase metals onto dust grains described in Section \ref{model_section}, also computed as the derivative of the dust-gas relation. The green dashed line shows the G/D measured from depletions.}
\label{gd_trends_fit}
\end{figure*}

\noindent We take the derivative of these functional form to derive the G/D: $G/D = d\Sigma_g/d\Sigma_d^{\mathrm{fit}}$. We bootstrap the errors on the fit using a simple Monte-Carlo simulation, by propagating the uncertainties on the $\Sigma_d$ measurement at each gas surface density bins into the fit and derivative. The resulting trends are shown in Figure \ref{gd_trends_fit} and are in excellent agreement with the G/D measured in \citet{RD2014} in the overlapping surface density range.

\subsection{Why does the dust abundance vary with surface density?}\label{model_section}

\indent The seeds of dust grains are formed in the cool and dense envelopes of evolved stars \citep{bladh2012} and Supernova remnants \citep{matsuura2011}. Recent modeling and observations have shown that present-day dust abundances can only be explained if dust grows in the ISM from accretion of gas-phase metals \citep{boyer12}. The most direct evidence of dust growth via accretion in the ISM comes from depletions \citep{jenkins09, tchernyshyov2015}, which indicate that the G/D decrease by a factor 3 between the most diffuse and densest sight-lines probed by UV spectroscopy of massive stars in the Milky Way, LMC, and SMC. Thus, dust grains grow in the dense ISM, and are then re-distributed to the more diffuse ISM when dense clouds are disrupted. In the diffuse ISM, dust grains are destroyed by interstellar shocks traveling at speeds 100 \kms \citep{jones94}. Dust grains are also consumed by star-formation. In the following, we develop a simple analytical model to explain the observed trends by dust growth in the ISM. \\
\indent Variations of dust abundance with density are a result of the equilibrium between production, mixing and destruction, and therefore of the differences between the corresponding timescales. Multi-phase (resolved) numerical simulations of dust evolution also find that the G/D can vary by factors of several between the diffuse and dense ISM \citep{bekki2013, zhukovska2016, mckinnon2016}. The timescale for dust formation and growth in the ISM corresponds to the collisional timescale between already existing dust grains and gas-phase metals, which in turn is inversely proportional to the gas density. Specifically, the dust growth timescale in the ISM, $\tau_{acc}$ is given in \citet{asano2013}:

\begin{multline}
\left ( \frac{\tau_{acc}}{yr} \right ) = 2\times 10^7 \times \\
\left ( \frac{<a>}{0.1 \mu m} \right ) \left (\frac{n_H}{100 \mathrm{cm}^{-3}} \right )^{-1} \left (\frac{T_d}{50 K} \right)^{-\frac{1}{2}} \left (\frac{Z}{0.02} \right )^{-1}
\label{timescale_eq}
\end{multline}

\noindent where $<a>$ is the mean dust grain radius (typically 0.1 $\mu$m), and the dust temperature in the LMC and SMC is $\sim$30K (see Figure \ref{plot_dust_params}). The metallicities of the LMC and SMC are $Z$ $=$ 0.5 $Z_{\odot}$ and  $Z$ $=$ 0.2 $Z_{\odot}$, respectively \citep{russell92}. The timescale for dust destruction by shocks is about 0.1 Gyr \citep[][and reference therein]{zhukovska2016}, while star formation only consumes dust grains on $>$2 Gyr timescales \citep{bigiel2008}. Therefore, interstellar shocks are the main dust destruction mechanism. \\
\indent Observationally,  elemental depletion measurements show that the dust-to-metal (D/M) ratio depends on metallicity and density. At the lowest densities ($F_*$ $\sim$ 0), about $f_0$ $=$ 20\% of metals are in the dust phase in the MW and LMC, while a lower fraction (5-10\%) of metals are locked in dust grains in the SMC. This number results from the timescales for dust growth, phase mixing, and dust destruction. \citet{zhukovska2008} solve the time differential equation governing the fraction of metals in dust as a function of time:

\begin{equation}
f_d = f_0  \frac{ \exp{\left (\frac{t}{\tau_{acc}}  \right )} }   {1-f_0 + f_0 \exp{\left ( \frac{t}{\tau_{acc}} \right ) } }
\label{gd_mod_eq}
\end{equation}

\noindent This equation also provides the density dependence of the dust-to-metal ratio (D/M) for a given time (through $\tau_{acc}$). Since abundance (M/G) are known in the LMC and SMC, we can derive the dust abundance (G/D). Therefore, we use this simple equation to model the dust abundance as a function of surface density. First, we assume that the mean density scales linearly with surface density, and that a mean density of $n_0$ $=$ 10 cm$^{-3}$ corresponds to a surface density of $\Sigma_0$ $=$ 25 \Msu pc$^{-2}$:

\begin{equation}
n_H = n_0 \frac{\Sigma_g}{\Sigma_0}
\end{equation}

\noindent  This implies that the medium has a mean density of 10 cm$^{-3}$ over a depth 100 pc, comparable to the gas disk thickness in the LMC \citep{elmegreen2001}. Based on this relation between density and surface density, we compute $f_d$ as a function of surface density $\Sigma_g$ from Equations \ref{timescale_eq} and \ref{gd_mod_eq} assuming $t=10^7$ yr, which is roughly a dynamical time for turbulence and the timescale for dissipation of molecular clouds due to star formation. This time should be a good estimate for the timescale on which phases mix. We assume $f_0$ $=$ 0.2 in the LMC and $f_0$ $=$ 0.05 in the SMC \citep{tchernyshyov2015}. Finally, we compute the trend in (G/D) vs $\Sigma_g$ via:

\begin{equation}
\left ( \frac{G}{D} \right ) = \left ( \frac{G}{D} \right )_0 \frac{f_0}{f_d}
\end{equation}

\noindent We take $(G/D)_0$ to be the gas-to-dust ratio in the most diffuse regions probed in our data set, corresponding to the maximum (G/D) in the trends shown in Figure \ref{gd_trends}.\\
\indent  The resulting trends of (G/D) as a function of $\Sigma_g$ are compared to the measurements in Figure \ref{gd_trends} (purple dashed line). Overall, we find a good agreement between this very simple model and the trends observed in the data. This is a good indication that dust indeed growth in the ISM by accretion of gas-phase metals, with density-dependent timescales consistent with simple accretion processes.

\subsection{Possible effects of optically thick \hi, CO-dark H$_2$,  emissivity variations, and resolution}

\indent \citet{RD2014} pointed out the degeneracy between the effects of G/D variations, molecular gas (H$_2$) dark in CO, optically thick \hi, and variations of the dust emissivity. Dust emissivity variations and dark gas (optically thick \his and CO-dark H$_2$) can mimic this effects of an increasing dust abundance with increasing surface density. \\
\indent When dust grains coagulate in the dense ISM, their emissivity (amount of emission per unit mass) increases. Since such variations cannot be accounted for in the modeling/fitting of the dust SED due to the absence of model or observational constraints for this effect, assuming a constant emissivity implies that the surface density of dust is over-estimated and the G/D under-estimated in the dense ISM. Since coagulation only occurs at high densities, the bias only occurs in the dense ISM, leading to an apparent decrease of G/D with increasing surface density. However, this effect only kicks in at gas surface densities where a large fraction of the gas is molecular. The \hi-H$_2$ transition occurs at $\Sigma_g$ $=$ 45 \Msu pc$^{-2}$ in the LMC and 110 \Msu pc$^{-2}$ in the SMC \citep{RD2014}, which is at the very high end of the surface densities probed here. The G/D variations observed here occur mostly below this threshold, and on scales larger than the typical scales of giant molecular clouds (GMCs). Hence, we do not expect coagulation and dust emissivity variations to be the cause of the trends observed here. \\
\indent Optically thick \his and CO-dark H$_2$ can lead to an apparent decrease of G/D with increasing surface density because this gas is essentially invisible to us. \citet{RD2014} demonstrated that the atomic-to-molecular transition occurs at $\Sigma_g$ $\sim$ 45 \Msu pc$^{-2}$ in the LMC and  $\Sigma_g$ $\sim$ 110 \Msu pc$^{-2}$ in the SMC. The trends observed here occur at surface densities lower than the \hi-H$_2$ transition and CO-dark H$_2$ is therefore unlikely to affect the trends. \\
\indent It is not clear how much optically thick \his contributes to the observed decrease of the G/D. However, this effect by itself could not explain the factor 4---10 variation in the G/D, or the required \his mass would be unreasonably high. Indeed, a study based on absorption 21 cm spectra in the SMC by \citet{dickey2000} demonstrated that the filling factor of the cold neutral phase is small ($<$15\% of the \his mass) in the SMC. They derived a correction of self-absorption, which takes the form $f_c = 1 + 0.667 \log(N($\hi$) - 21.4)$ for $N($\hi$) > 21.4$ and $f_c$ $=$ 1 otherwise. At the highest column densities probed in this study ($\Sigma_g$ $=$ 100 corresponding to $N($\hi$)$ $=$ $10^{22}$ cm$^{-2}$), this correction factor is 1.4. Therefore, optically thick \his in the SMC does not come close to explain the decrease in G/D seen in Figure \ref{gd_trends}. No constraints on the mass of optically thick \his emission is available in the LMC, but it is unlikely that it would explain a factor 3 decrease in the observed G/D given the magnitude of the effect in the SMC.  \\
\indent Dust emission depends non-linearly on the dust temperature. As a result, temperature mixing in the beam can result in an underestimation of the true dust surface density is the resolution is too coarse to resolve the cold ISM \citep{galliano2011}.  However, \citet{galliano2011} have shown that, in the LMC and SMC, the dust surface density estimate starts to suffer from this effect (at the $>$ 10\% level) for resolutions coarser than 100 pc (see their Figure 6). The resolution of our maps is 75 pc (LMC) and 90 pc (SMC), and the dust surface density estimate should therefore be close to what would be derived at higher resolution. Indeed, our derived dust surface densities and mass values agree well with \citet{gordon2014} and \citet{RD2014}, who estimated the dust content of the LMC and SMC at 10 pc resolution (\citet{RD2014} convolved the dust surface density to 15 pc (LMC) and 45 pc (SMC), but the dust SED fitting was performed at the native Herschel resolution of 36"). The relatively low resolution of the PLANCK and IRAS data therefore does not play a role in the observed G/D trends.

\subsection{Implications}

\indent The variations of the dust abundance observed in this analysis have important implications in several astrophysical fields. First, dust surface density maps or dust masses are frequently used as gas tracers, via the assumption of a constant G/D, both at low redshift \citep[e.g., ][]{israel1997, leroy2007, leroy2009, bolatto2011, jameson2016} and high redshift \citep{rowlands2012}. Non-linear variations of G/D with metallicity and large variations of the G/D with surface density could substantially bias the gas and dust mass estimates of galaxies.\\
\indent Second, the results presented here provide some observational constraints needed by recent chemical evolution models in order to narrow down timescales for dust formation and destruction in different regions of galaxies (star-forming, diffuse, boundary with CGM). In the future, we will compare our results with some of these models in order to advance the field of dust evolution. 
\indent Third, numerical models of the ISM \citep[e.g., ][]{glover2010, glover2011, smith2014} typically assume a single G/D, scaled linearly with metallicity, and constant dust properties. Since dust plays a key role in the radiative transfer in the ISM, which in turns determines the structure and composition of the different ISM phases, it is crucial to incorporate variations in the dust abundance and properties. Our results can serve as a prescription for these numerical models.

\section{Conclusion}\label{conclusion_section}

\indent We investigate the variations of the G/D and dust properties as a function of gas surface density using a stacking analysis of \his 21 cm, CO, PLANCK and IRAS data in the LMC and SMC. \\
\indent We account for systematics by correcting the \his 21 cm maps for stray-light and by carefully estimating and subtracting foreground cirrus emission from the Milky Way, using two methods (a linear and a quadratic fit to the relation between galactic \his and dust emission). We stack the LMC and SMC FIR emission observed in the IRAS 100, PLANCK 350, 550, and 850 $\mu$m bands in bins of gas surface density, and model the resulting stacked dust SED in each gas surface density interval using a modified black body fit performed within a probabilistic framework which accounts for background uncertainties, measurement, and calibration errors. We thus derive the dust surface density, temperature and spectral emissivity index as a function of gas surface density. \\
\indent We find no significant variation in the dust temperature and spectral emissivity index with gas surface density on the scales probed with our all-sky surveys (75 pc in the LMC, 90 pc in the SMC). The dust temperature lies in the range 25---30 K in both galaxies and the spectral emissivity index is constant at 0.8---1 in the SMC and 1.2---1.4 in the LMC, indicating that dust grains may be pre-dominantly composed of amorphous carbon. \\
\indent The dust surface density correlates non-linearly with gas surface density, resulting a G/D decreasing by factors of 4 (LMC) to 10 (SMC) between with the observed range of gas surface density ($\sim$ 10---100 \Msu pc$^{-2}$). The observed trend is consistent with depletion measurements in the LMC. In the SMC, the slopes of the G/D vs gas surface density relation are consistent between depletions and our measurement, but the value of the measured G/D are discrepant by an order of magnitude, which could be explained by observational biases in the UV spectroscopic absorption \his measurements linked to the complex velocity structure of the SMC. \\
\indent The G/D measurements in the LMC and SMC indicate a non-linear relation between dust abundance and metallicity, with the dust content of the SMC being a factor of 4 lower than what a linear scaling with metallicity would yield. This is consistent with the conclusions of recent chemical evolution models, which point out that dust growth in the ISM becomes too inefficient at sub-SMC metallicity to counteract metal loss by galactic-scale outflows, and that the dust content of such low metallicity galaxies should be dominated by dust input by evolved stars, resulting in a non-linear relation between G/D and metallicity.\\
\indent We present a analytical model, simplified from numerical chemical evolution models, which explains the observed relation between dust and gas surface densities by accretion of gas-phase metals onto dust grains with a density-dependent timescale.  This model reproduces the observations, indicating that dust growth and destruction in the ISM plays a major role in the evolution of galaxies.

\bibliographystyle{/users/duval/stsci_research/bibtex/apj11}
\bibliography{/Users/duval/stsci_research/biblio_all}

\acknowledgments{Julia Roman-Duval acknowledges support from the European Space Agency. We thank the anonymous referee for making useful suggestions that improved the quality of the paper.

}

\end{document}